\documentclass[journal]{IEEEtran}
\IEEEoverridecommandlockouts
\usepackage{cite}
\usepackage{amsmath,amssymb,amsfonts, amsthm}
\usepackage{verbatim}
\usepackage[acronym]{glossaries}
\usepackage{algorithmic}
\usepackage{graphicx}
\usepackage{textcomp}
\usepackage[normalem]{ulem}
\def\BibTeX{{\rm B\kern-.05em{\sc i\kern-.025em b}\kern-.08em
    T\kern-.1667em\lower.7ex\hbox{E}\kern-.125emX}}

\usepackage{tikz}
\usetikzlibrary{shapes, arrows, positioning, plotmarks, patterns}    
\usepackage{pgfplots}
\pgfplotsset{
compat=1.18,
yminorticks=false,
xminorticks=false,
tick style={color=black},
every tick label/.append style={font=\footnotesize},
tick label style={font=\small},
grid style={line width=.1pt, draw=gray!20},
major grid style={line width=.1pt,draw=gray!20},
}
\usepgfplotslibrary{colorbrewer, groupplots, fillbetween}
\usetikzlibrary{decorations.pathreplacing}

\usepackage[caption=false]{subfig}

\newtheorem{theorem}{Theorem}
\newtheorem{lemma}{Lemma}

\newcommand{\Tau}{\mathcal{T}}
\newcommand{\E}[1]{\mathbb{E}\left[ #1 \right]} 
\newcommand{\N}{\mathcal{N}}
\newcommand{\G}{\mathrm{Gamma}}
\newcommand{\bm}[1]{\pmb{ #1 }}

\newcounter{Prob}
\newenvironment{problemeq}
  {\stepcounter{Prob}%
    \addtocounter{equation}{-1}%
    \equation}
  {\endequation}

\newcommand{\AM}[1]{#1}
\newcommand{\FC}[1]{#1}
\newcommand{\PT}[1]{#1}
\newcommand{\AZ}[1]{#1}
\newcommand{\pp}[1]{\textcolor{magenta}{#1}}
\newcommand{\rev}[1]{#1}

\definecolor{color1}{RGB}{240, 249, 33}
\definecolor{color2}{RGB}{253,180,47}
\definecolor{color3}{RGB}{237,121,83}
\definecolor{color4}{RGB}{204,71,120}
\definecolor{color5}{RGB}{156, 23, 158}
\definecolor{color6}{RGB}{92,1,166}
\definecolor{color7}{RGB}{13,8,135}

\begin{document}

\newacronym{ts}{TS}{time sensitive}
\newacronym{aoi}{AoI}{age-of-information}
\newacronym{voi}{VoI}{value-of-information}
\newacronym{qaoi}{QAoI}{query age-of-information}
\newacronym{bs}{BS}{base station}
\newacronym{harq}{HARQ}{hybrid automated repeat request}
\newacronym{ee}{EE}{energy efficiency}
\newacronym{adc}{ADC}{analog-to-digital converter}
\newacronym{urllc}{URLLC}{ultra reliable low latency communication}
\newacronym{iot}{IoT}{internet of things}
\newacronym{pdf}{PDF}{probability density function}
\newacronym{cdf}{CDF}{cumulative distribution function}
\newacronym{kpi}{KPI}{key performance indicator}
\newacronym{snr}{SNR}{signal-to-noise ratio}
\newacronym{arq}{ARQ}{automatic repeat request}
\newacronym{iid}{i.i.d}{independent and identically distributed}
\newacronym{5g}{$\textrm{5}$G}{fifth-generation}
\newacronym{6g}{$\textrm{6}$G}{sixth-generation}
\newacronym{cpu}{CPU}{central processing unit}
\newacronym{wsn}{WSN}{wireless sensor networks}
\newacronym{spa}{SPA}{Saddle Point Approximation}
\newacronym{goc}{GoC}{Goal-oriented Communication}
\newacronym{cscg}{CSCG}{circularly symmetric complex Gaussian}
\newacronym{clt}{CLT}{central limit theorem}
\newacronym{paoi}{PAoI}{peak AoI}
\newacronym{csi}{CSI}{channel state information}

\title{End-to-end Modeling and Optimization of 
Timing and Energy in Wireless IoT Systems

\thanks{P. Talli (corresponding author -- pietro.talli@phd.unipd.it), F. Chiariotti (federico.chiariotti@unipd.it), and A. Zanella (andrea.zanella@unipd.it) are with the Department of Information Engineering, University of Padova, Italy. A. Mishra (anmi@es.aau.dk), I. Leyva-Mayorga (ilm@es.aau.dk), and P. Popovski (petarp@es.aau.dk) are with the Department of Electronic Systems, Aalborg University, Denmark. This work was supported by the European Union under the Italian National Recovery and Resilience Plan of NextGenerationEU, under the partnership on ``Telecommunications of the Future'' (PE0000001 - program
``RESTART'') and, in part, by the Velux Foundation, Denmark, through the Villum Investigator Grant WATER, nr. 37793. Anup Mishra is supported by the Horizon Europe Marie Sk{\l}odowska-Curie Actions (MSCA) Postdoctoral Fellowships under the project {FOCUS-6G} (Grant Agreement No.~101202623).}
}

\author{Pietro Talli,~\IEEEmembership{Student Member,~IEEE,} Anup Mishra,~\IEEEmembership{Member,~IEEE,} Federico Chiariotti,~\IEEEmembership{Senior Member,~IEEE,}\\ Israel Leyva-Mayorga,~\IEEEmembership{Member,~IEEE,} Andrea Zanella,~\IEEEmembership{Senior Member,~IEEE,} and Petar Popovski,~\IEEEmembership{Fellow,~IEEE}\vspace{-0.4cm}}

\maketitle

\begin{abstract}
\AZ{With the advent of edge computing, data generated by end devices can be pre-processed before transmission, possibly saving transmission time and energy. On the other hand,  data processing  itself incurs latency and energy consumption, depending on the complexity of the computing operations and the speed of the processor. The energy-latency-reliability profile resulting from the concatenation of pre-processing operations (specifically, data compression) and data transmission is particularly relevant in wireless communication services, whose requirements may change dramatically with the application domain. In this paper, we study this multi-dimensional optimization problem, introducing a simple model to investigate the tradeoff among end-to-end latency, reliability, and energy consumption when considering compression and communication operations in a constrained wireless device. We then study the Pareto fronts of the energy-latency trade-off, considering data compression ratio and device processing speed as key design variables. Our results show that the energy costs grows exponentially with the reduction of the end-to-end latency, so that considerable energy saving can be obtained by slightly relaxing the latency requirements of applications.} These findings challenge conventional rigid communication latency targets, advocating instead for application-specific end-to-end latency budgets that account for computational and transmission overhead.
\end{abstract}


\section{Introduction}
Sensor-based wireless data collection is an essential service of modern communication networks, 
supporting an ever-widening set of applications, from long-term environmental monitoring to virtual reality and industrial automation~\cite{marjani2017big,YICK20082292}. 
Notably, as the variety of sensing scenarios expands, sensor nodes evolve to incorporate enhanced computing capabilities. These capabilities can be leveraged to compress the sensed data before transmission, potentially reducing delay and energy costs of transmission. On the other hand, data processing itself incurs costs in terms of delay and energy consumption. \AM{Therefore, the typical \glspl{kpi} of wireless networks, namely latency, reliability, and \gls{ee}, are jointly influenced by decisions made in both computation and communication stages\cite{varghese2021survey,Israel@Power_Freq,Shao@Comm_Comp,Chen@Tradoff}.}
\par Higher (lossless) compression ratios generally require more processing cycles and, consequently, longer processing time. This can be reduced by increasing the processing speed, at the cost of a higher computing power.  On the other hand, compression reduces the amount of data to be transmitted, thus saving transmission time and potentially energy for a certain level of reliability\cite{barr@energy}. 
In parallel, communication time can be significantly reduced by increasing the transmission data rate. However, \PT{under a fixed bandwidth,} this approach requires higher transmission power to reach a \gls{snr} sufficient to ensure reliable data transmission despite the increased speed.

\AM{While the general trade-offs between energy, latency and reliability are well known at the level of individual operations} \cite{Chen@Tradoff,Avranas@EL_Tradoff,varghese2021survey}, their interplay becomes more nuanced in the context of an end-to-end chain of operations within a wireless sensor network possessing computational capabilities. This complexity arises from the coupled effects of compression and transmission decisions: while increasing the compression ratio makes transmission faster and more efficient, it also has a computational cost that affects both the latency and the sensor's energy budget.
Consequently, there is a growing need for improved modeling and understanding of the interplay between system design parameters and device-level constraints. By simultaneously modeling computation and communication, we can gain deeper insights into optimal resource allocation strategies. 
This approach enables us to better balance and optimize energy consumption, latency, and reliability across the entire system.
\PT{While our own recent works~\cite{suman2023statistical,li2025unified} considered the end-to-end latency, including both computational and communication components, the trade-off between the latency and the overall energy consumption, considering all steps of the process, is still relatively unexplored, and there is limited work on their joint optimization.}

\subsection{Contribution}
\label{subsec:contribution}
\PT{In order to address this gap, we present a constrained optimization framework that enables joint computation and communication systems to select strict energy requirements, minimizing the worst-case end-to-end transmission latency in an \emph{energy-constrained scenario}.} To this end, we consider an  open-loop system where a sensor transmits uplink data to a \gls{bs}. The computational aspect involves compression at the sensor and decompression at the \gls{bs}.
\PT{We consider a tight upper bound of the latency, enabling system designers to select the required reliability quantile. \rev{For a fixed physical-layer configuration, we formulate an energy-constrained latency optimization problem, adapting the compression ratio and processing frequency under a total energy budget.} Furthermore, we consider the dual \emph{time-constrained scenario}, in which there is a strict constraint on timeliness, expressed as a requirement on the worst-case \gls{qaoi}~\cite{Federico@QAoI}, and the goal is to minimize the energy required to satisfy it.} \rev{As in the energy-constrained scenario, the optimization is conditional on fixed lower-layer communication parameters.}

We focus on lossless compression, so that the fidelity of the reconstructed data is preserved, and accuracy is not a concern. 
By treating the data compression ratio and processing speed as design variables, we identify system operating points and demonstrate that modest relaxations in latency requirements can lead to substantial energy savings. Leveraging modeling approximations developed in our prior works~\cite{suman2023statistical,li2025unified}, \PT{which characterized the end-to-end latency of joint communication-computation scenarios,} we extend the analysis \PT{by modeling the energy of} both computation and communication across varying reliability levels \PT{and using it in our optimization}. This provides a more comprehensive perspective on system-level energy–latency optimization, which covers a gap in the current literature, as discussed in the next section.

The major contribution of this paper are the following:
\begin{itemize}

\item \PT{We extend the end-to-end latency characterization from\cite{suman2023statistical,li2025unified} to consider the energy cost of joint computation-communication operations, capturing compression at the sensor, decompression at the \gls{bs}, and packet-based transmission with reliability constraints, enabling a unified analysis of processing and communication operations.}

\item \PT{We formulate an energy-constrained latency optimization problem, }where the goal is to minimize the $\varrho$-quantile of the end-to-end latency under a total energy budget. \PT{By considering a tight upper bound of the latency quantile, we prove the convexity of the problem under mild conditions, showing that it can be solved efficiently.}

\item We formulate a complementary {time-constrained optimization problem}, relevant for scheduled access scenarios governed by \gls{qaoi} constraints. In this setting, we minimize energy consumption while meeting a fixed \gls{qaoi} budget and target reliability. While this problem is more complex, we derive the optimal allocation of compression time and transmission time to jointly satisfy these constraints.

\item Through Pareto front analysis, we quantify the trade-offs between latency and energy consumption across varying reliability levels. Our results show that modest relaxations in latency constraints can lead to substantial energy savings. We also demonstrate how system-level parameters, such as compression ratio and processing speed, can be jointly tuned to optimize both latency and energy consumption under different operational constraints.
\end{itemize}

The rest of this paper is organized as follows. \AZ{Sec.~\ref{sec:relatedwork} provides a quick overview of the state of the art of energy-latency and reliability analysis in wireless networks, highlighting the gaps covered by this work.} Sec.~\ref{sec:sys} provides detailed explanations of the analytical models for compression, decompression, and transmission that we use throughout the paper. In Sec.~\ref{sec:RCS_Section} we describe and analyze the two considered scenarios. For each of them, we provide an analytical characterization of the latency, energy consumption and reliability. Moreover, we define an optimization problem and discuss solutions and optimization strategies. Sec.~\ref{sec:results} then presents the results of simulations with realistic parameters, which show how the proposed trade-offs can arise in real-world devices and networks. \AZ{Finally, Sec.~\ref{sec:conclusion} summarizes the key findings and proposing potential directions for future research.}
\subsection{Notation}
\AM{Matrices are denoted by boldface uppercase letters, column vectors are denoted by boldface lowercase letters, and scalars are denoted by standard letters. $\mathbb{E}_{X}[Y]$ denotes the expectation of $Y$ with respect to the random variable $X$. $\mathbb{C}^{M\times N}$ and $\mathbb{R}^{M\times N}$ denote the sets of all $M \times N$ matrices with complex- and real-valued entries, respectively. The \gls{cscg} distribution with mean $\mu$ and variance $\sigma^{2}$ is denoted by $\mathcal{CN}(\mu,\sigma^{2})$. The Big-O notation $\mathcal{O}(f(n))$ describes the asymptotic upper bound of an algorithm's computational complexity. The floor and ceiling functions are denoted by $\lfloor x \rfloor$ and $\lceil x \rceil$, representing the greatest integer less than or equal to $x$ and the smallest integer greater than or equal to $x$, respectively.}
\section{Related Work}\label{sec:relatedwork}
\AM{This work builds on the well-known trade-off between source coding and channel coding~\cite{SourceVsChannelCode,Hag@SourceControlled_Channel}}. 
\AM{In the context of communication systems, the energy–latency trade-off has received considerable attention in the literature~\cite{Berry@Delay,Avranas@EL_Tradoff,Mirza@EL}. For example, \cite{Berry@Delay} investigates the delay and power consumption trade-off incorporating buffer constraints. In \cite{Avranas@EL_Tradoff}, the authors explore energy-latency trade-offs in \gls{urllc} systems by optimizing retransmissions and blocklength. Similarly,~\cite{Mirza@EL} develops a scheduling algorithm grounded in finite blocklength theory and derives energy–delay bounds.}

\PT{The energy-latency trade-off has been widely examined in the context of Edge computing~\cite{li2025unified} and task offloading~\cite{ballotta2020computation}: in particular, recent works have focused on the inference performance of machine learning systems~\cite{shao2021communication,cong2023latency}. Naturally, learning-based systems face significant challenges in terms of energy efficiency~\cite{li2017fundamental}, and the joint consideration of computing and communication operations leads to fundamentally different trade-offs~\cite{zhang2022dependent}. However, another aspect has recently captured the attention of the research community: while the significant energy costs of learning operations need to be considered even in the Cloud, sensors and battery-powered devices face stringent energy constraints for far simpler operations. The reliability and latency requirements of 5G and 6G communications may strain the energy budget of these devices~\cite{elgarhy2024energy}, and are a prime target for improvement and optimization~\cite{bisio2024shm}. The recent literature has extensively analyzed the effect of communication on the lifetime of \gls{iot} nodes, modeling it accurately in a variety of scenarios~\cite{Mao@Freq_Power}, but similar efforts that consider the compression of complex data are currently lacking. }

\PT{In fact, compression optimization has been considered at the architectural level~\cite{burd1996processor,de2013energy}, but only sparse recent efforts~\cite{zeng2023parallelizing} have considered the specifics of \gls{iot} nodes, such as, e.g., the difficulty of parallelizing tasks. Our previous works~\cite{suman2023statistical,li2025unified} primarily focused on end-to-end latency modeling/optimization across compression and transmission, while~\cite{Israel@Power_Freq} addressed energy minimization under real-time processing/communication constraints in a satellite edge-computing setting. However, these works failed to provide an end-to-end characterization of the energy--latency trade-off under reliability constraints. Similarly, \cite{ballotta2020computation} jointly optimizes computation and communication for real-time monitoring of dynamic systems, but with limited energy considerations.}

\PT{The literature on \gls{aoi}~\cite{huang2021age} has examined the effect of energy constraints in different scenarios, often considering multi-user scenarios~\cite{das2025analysis} and random access optimization~\cite{tang2020minimizing}, but it generally focuses on access and sampling strategies over time, even when considering data compression~\cite{hu2022age} as part of the system model, rather than on the compression and computation of an individual update with specific timing requirements.} Other works which considered latency-accuracy trade-offs are in \cite{Israel@Power_Freq,  li2026task}. Another recent work~\cite{long2025energy} poses the energy-latency trade-off for vehicular networks, considering the offloading of computing tasks from computers to an edge-enabled \gls{bs}. Our work differs from these studies by considering generic data types and by explicitly modelling the energy consumed by both computation and communication. In particular, \cite{li2026task} uses a learning-based algorithm to determine context-dependent offloading strategies, whereas our work develops an analytical model of the resulting energy--latency--reliability trade-off.

In summary, most of existing works in the literature are constrained to predefined latency-reliability requirements, \PT{without much flexibility for the actual timing requirements of the applications}~\cite{Time@Petar}\PT{, and considering latency and energy efficiency as entirely separate}. Moreover, in many network scenarios, it remains an open question whether it is more beneficial to invest energy in compressing data sources to reduce communication overhead, or to allocate more energy to channel coding to enhance reliability\AM{~\cite{Shao@Comm_Comp,li2025unified}}. \PT{Interestingly, one of the few works to explicitly consider both energy and latency considers complexity at the decoder~\cite{celebi2021multi}, targeting a downlink scenario, while we aim at an uplink scenario in which the battery-powered node needs to compress sensory data.}
\PT{Hence, the joint optimization of energy and latency in \gls{iot} nodes is an open problem, which we tackle in this work by drawing on state-of-the-art models from the literature and combining them into a dual objective optimization problem.}

\section{System Model}\label{sec:sys}

In this section, we present the energy and latency models for an open-loop system, where a wireless sensor is transmitting uplink data to the \gls{bs}. We define a \textit{Computational Model} which entails the compression at the sensor and decompression at the \gls{bs}, and a \textit{Communication model} that captures the data transmission over the wireless channel. 
We characterize each model in terms of latency and energy requirements, and then analyze their joint optimization in the following section.

\subsection{Computational Model}\label{sec:comp_model}
\AM{For analytical simplicity, we focus on lossless compression in this work.\footnote{\AM{Incorporating lossy compression would introduce the additional dimension of accuracy. This will be explored in future work to examine the trade-offs between energy, latency, reliability, and accuracy.}}} We consider data to come in \emph{blocks}, which are the basic unit that needs to be compressed and transmitted. The compression ratio $Q\in [1,Q_{\mathrm{max}}]$ is defined as the ratio \FC{between the size of the original data block and the size of the corresponding compressed output}. \AZ{The maximum value $Q_{\mathrm{max}}$ corresponds to the maximum possible data reduction in lossless compression.} 

\FC{Common lossless compression algorithms use dictionary-based methods such as Lempel-Ziv~\cite{ziv1977universal}, which require scanning through the content multiple times to further reduce the output size.} \AZ{Therefore, larger compression ratios generally require  a greater computational effort\cite{li2018wirelessly}.} \PT{The most common implementations of lossless algorithms are LZ77, DEFLATE and LZMA, which represent the state of the art in many applications when transmitting feedback data such as logs and general different formats.} Quantitatively, \AM{computational complexity} can be expressed in terms of number of required cycles of the \gls{cpu} for bit of processed data. \PT{This number can be modeled as a Gamma-distributed random variable $X_c \sim \G(\kappa, \beta)$,
characterized by its shape parameter $\kappa$ and scale parameter $\beta$~\cite{jovsilo2018selfish,han2019offloading}. The complexity of algorithms such as LZ77 has been widely studied and experimentally validated~\cite{arroyuelo2021engineering}, and the use of a Gamma distribution can be intuitively explained by considering the total time of repeated events, which in this case represent re-occurring sequences. The use of a Gamma distribution can then be reasonably extended to other compression algorithms and is not intended to be specifically tailored to LZ77.
} \AZ{Following\cite{li2018wirelessly}, the relation between the compression ratio $Q$  and the mean \AZ{computational} complexity $\E{X_c}$ can be modeled as:}
\begin{equation}\label{eq:mean_complexity}
    \E{X_c} = e^{\psi Q} - e^{\psi},
\end{equation}
where $\psi>0$ is a constant that depends on the compression algorithm used~\cite{li2018wirelessly,wang2020joint}. \PT{The same model has been proposed to describe compression time of other algorithms such as JPEG \cite{Israel@Power_Freq}. However, considering lossy compression algorithms introduces another dimension to the problem, i.e., the accuracy of the transmitted data, which goes beyond the scope of this work}. \AM{From  \eqref{eq:mean_complexity}, it can be discerned that} the expected number of operations increases exponentially with $Q$. \PT{We recall that the expected value of the Gamma distribution is $\kappa\beta$, which gives the following relationship between $\beta$ and $Q$
\begin{equation}
    \beta = \frac{e^{\psi Q} - e^{\psi}}{\kappa}.
\end{equation}}
The compression time $T_c$ for a data block of $D$ bits is directly proportional to $X_{c}$, and can be expressed as~\cite{varghese2021survey}: 
\begin{equation}\label{eq:compression_time}
    T_{c} = f_c^{-1}X_{c}\,D,
\end{equation}
where $f_{c}$ is the \gls{cpu}-cycle frequency of the device\cite{li2018wirelessly}. From \eqref{eq:mean_complexity} and \eqref{eq:compression_time}, it follows that $T_c \sim \G(\kappa, {\beta D}{f_{{c}}^{-1}})$. 

\FC{ The number of operations $X_d$ necessary for decompression is generally lower than $X_c$ and can be modeled as a deterministic fraction $\zeta \in (0,1]$ of $X_c$~\cite{kothiyal2009energy,burrows1992line,zhang2020efficient}. We also note that decompression is performed at the \gls{bs}, which typically operates under less stringent energy constraints and has greater computational capacity, further decreasing the impact of $T_d$ on the overall latency.} 
Some implementations of LZ77 and LZMA can support streaming compression by compressing chunks of data over a fixed window size. However, in order to be able to accurately control the compression ratio $Q$, the window size should be typically larger than the data size $D$ considered in this work (in the order of $500$~kB, although different data sizes are considered in one simulation). This is why, in both scenarios, we require compression of all the data to be completed before the first packet can be sent. In addition, compression of a single chunk cannot be easily parallelized. While multithreading can yield some minor performance gain even in single chunk compression, the core part of the compression algorithm, represented by the entropy coding, has to be performed sequentially.

\par \AZ{The \gls{cpu}-cycle frequency $f_c$ also affects the average power consumed during the compression operation. In fact, according to  \cite{Israel@Power_Freq,Mao@Freq_Power}, the power absorbed per \gls{cpu}-cycle grows cubically with the \gls{cpu} clock. Therefore, denoting by $P_{{s},\mathrm{max}}$ the maximum processing power at the sensor, and by $f_{c,\mathrm{max}}$ the maximum \gls{cpu}-cycle frequency, the average power consumed per \gls{cpu} cycle can be estimated as  \cite{Israel@Power_Freq}: }
\begin{equation}\label{eq:computation_Power_freq}
    P_{c}(f_{{c}}) = P_{s,\mathrm{max}} \left(\frac{f_{{c}}}{f_{{c},\mathrm{max}}}\right)^3. 
\end{equation}
Using  \eqref{eq:computation_Power_freq}, the average energy consumption for the compression operation can be calculated as\cite{suman2023statistical}:

\begin{equation}
   E_{{c}}(f_{{c}},Q)     =P_{{c}}(f_{{c}})\E{T_{c}}
     = P_{s,\mathrm{max}} \frac{(e^{\psi Q}-e^\psi)Df_c^2}{(f_{c,\rm{max}})^3}.
\end{equation}
The energy for decompression at the \gls{bs} depends on its processing power, but follows the same model. 

In the processing chain considered here, the compression window is assumed to encompass the entire $D$-bit data block, so compression is completed before transmission begins. The resulting bitstream, containing approximately $D/Q$ bits, is then passed to the communication stage, where it is segmented into packets for transmission over the wireless channel described in Sec.\ref{sec:comm_model}.

\subsection{Communication model}\label{sec:comm_model}
In order to be transmitted over the channel, the compressed data block is split into \emph{packets}. 
We consider a packet to be composed of $N_b$ symbols, with a symbol time $B^{-1}$, where $B$ is the channel bandwidth.
We consider a linear channel model such that, given the input signal $\bm{x}[n]$ and blocklength $N_b$, the received signal is
\begin{equation}
    \bm{y}[n] = h[n]\bm{x}[n] + \bm{z}[n],
\end{equation} 
where $\bm{z}[n] \sim \mathcal{C}\mathcal{N}(0,N_0\bm{I}_{N_b})$ and $\frac{N_0}{2}$ is the noise power spectral density.
The channel gain $g = |h|^2$ is the product of a constant path loss value $\bar{g}$ and a fading parameter $f$. We denote the uplink transmission power as $P_{\textrm{tx}}$ and bandwidth as $B$. The \gls{snr} at the \gls{bs} for a transmitted packet can be expressed as\cite{suman2023statistical}:
\begin{equation}\label{eq:snr_tx}
    \gamma (P_{\mathrm{tx}}, B, d) = \frac{K_0 P_{\mathrm{tx}}}{d^\ell N_0 B}g = \gamma_0 (P_{\mathrm{tx}}, B, d) f,
\end{equation}
where $K_0$ is the Friis equation parameter, $\ell$ is the path loss exponent, $N_0$ is the noise power spectral density, $d$ is the  sensor-\gls{bs} distance, and $\gamma_0$ is the average \gls{snr}\AM{\cite{tse2005fundamentals,suman2023statistical}}. 

We then consider two cases, which we name \emph{slow fading} and \emph{fast fading}. In the former case, the wireless channel is constant for the whole duration of the processing and communication operations, i.e., its coherence time is larger than the time needed to compress and transmit the full data block, or $T_{\text{coh}}>T_{\text{tx}}+T_c$~\cite{tavildar2006approximately}. This is the case for, e.g., line of sight-dominant channels with Rician fading~\cite{ozdogan2019massive}.
Following the derivation from \cite{li2024reliability}, which is based on the finite blocklength analysis from~\cite{tan2015third} and \cite{polyanskiy2010channel}, we can write that the maximum payload size $n_p(\varepsilon)$ for a target error probability $\varepsilon$, expressed in terms of the number of payload bits each codeword can encode, is 
\begin{equation}
\begin{aligned}
\label{eq:fb_rate}
    n_p(\varepsilon) \simeq &N_b\log_2(1+\gamma)+ \frac{\log_2N_b}{2}\\& - \sqrt{N_b \left( 1-\frac{1}{(1+\gamma)^2} \right)}Q^{-1}(\varepsilon)\log_2e,
\end{aligned}
\end{equation} 
where $\gamma$ is the \gls{snr}, which can be estimated by sending pilots before the compression and communication operations begin and is constant throughout the transmission. Knowing $\gamma$ is necessary for the sender to adapt the coding rate for the specific Gaussian channel instantiated by that $\gamma$.

On the other hand, in the fast fading case we consider a Rayleigh block fading model, in which the coherence time is larger than the time necessary to transmit a single packet, but smaller than the time necessary for the full transmission. Each packet then has an independent exponentially distributed fading gain $ f \sim \text{Exp}(1)$~\cite{tse2005fundamentals}. In this case, the fading cannot be predicted in advance, but we can still compute a transmission rate that ensures the target $\varepsilon$. We use the total probability to compute the expected error probability for a given payload size:
\begin{equation}
    \varepsilon(n_p) = \int_{0}^\infty e^{-f}  Q\left(\frac{N_b\log_2(1+f\gamma_0)+ \frac{\log_2N_b}{2} -n_p}{ \sqrt{N_b \left( 1-\frac{1}{(1+f\gamma_0)^2} \right)}\log_2e}\right) df.
\end{equation}
Given the target packet error probability $\varepsilon^*$, the maximum allowed payload size is then
\begin{equation}
n_{p,u}(\varepsilon^*)=\sup\left\{n\in\mathbb{N}:\varepsilon(n)\leq\varepsilon^*\right\}.
\end{equation}
We can then find this value numerically.

\PT{We then derive the packet time $t_p$ as 
\begin{equation}\label{eq:transmit_time}
    t_p =\frac{N_b}{B}.
\end{equation}}
In the fast fading case, we consider a guard interval to ensure that the fading is not correlated, using $T_{\text{coh}}$ as the minimum time necessary for each packet.
\AZ{We consider a persistent \gls{arq} scheme, wherein a packet is automatically retransmitted until it is correctly decoded and acknowledged by the receiver. For simplicity, we assume an ideal feedback channel. This assumption does not compromise the accuracy of our latency-energy model, as potential downlink errors could be accounted for in the link budget of the uplink transmission.}  \rev{We consider ARQ without combining across retransmissions. The use of hybrid \gls{arq}, which is considered in the 5G and 6G specifications~\cite{ts38321}, may further improve latency by requiring fewer retransmissions. However, we consider this to be outside the scope of this work since, even though the specific implemented mechanism would affect the total transmission time for each packet, it would have a limited impact on the design of the proposed framework.}

\AM{Subsequently,} we consider the standard definition of \gls{ee} $\eta$ given in \cite{Emil@EE} as:
\begin{equation}\label{eq:energy_efficiency_1} 
    \eta\mathrm{\ [bit/Joule]} = \frac{\mathrm{Payload\ data\ rate\ [bit/s]}}{\mathrm{Power \ consumption\ [Joule/s]}}.
\end{equation}
The payload data rate in the numerator is $R(\varepsilon)=\frac{n_p(\varepsilon)}{t_p}$.
Note that the denominator of  \eqref{eq:energy_efficiency_1} accounts for both the transmission power and circuit power~\cite{Emil@EE} that, in turn, accounts for the power consumption of the \gls{adc} for sampling at a given bandwidth $B$, as well as the power required for the channel encoding and decoding processes. While the former can be expressed as $\nu B$, the latter is proportional to the data rate and can be calculated as $\lambda R(\varepsilon)$\cite{Emil@EE}. Both $\nu$ and $\lambda$ are hardware-specific parameters. We then obtain the \gls{ee} as: 
\begin{equation}\label{eq:energy_efficiency_2}
    \eta(\varepsilon) = \frac{R(\varepsilon)}{P_{\mathrm{tx}} + \nu B + \lambda R(\varepsilon)}.
\end{equation}
We use  \eqref{eq:energy_efficiency_2} in the following section to compute the power required for data transmission in the respective target scenarios, from which we can obtain the energy consumed during the transmission step.

\section{Optimization under Resource Constraints}\label{sec:RCS_Section}
In this section, we analyze two representative resource-constrained  scenarios typical of sensor-based wireless systems and compute their overall latency and energy consumption for varying reliability levels. The proposed framework jointly selects the compression ratio $Q$, the \gls{cpu} frequency $f_c$, and, in the time-constrained scenario, the allocation factor $\alpha$, to satisfy end-to-end energy, latency, and reliability requirements. These variables couple the computation and communication stages through the processing cost, compressed payload size, and allocation of time between compression and transmission. The optimization is performed conditional on a fixed physical-layer configuration, including $P_{\mathrm{tx}}$, $B$, $N_{\mathrm b}$, and the target packet error probability, which are treated as parameters of the underlying communication layer and remain unchanged within each optimization instance.

Note that, as many applications are sensitive to latency peaks, rather than average values, we consider the $\varrho$-quantile of the latency, as we are interested in worst-case performance. Conversely, we only consider average energy consumption, as batteries smooth out the instantaneous variations in used energy and, therefore, performance only depends on the mean value.
The two formulations are complementary, and the reliability parameter $\varrho$ has been introduced in both scenarios with the same interpretation: it denotes the probability of successful data block delivery within the specified latency.  In the energy-constrained scenario, the objective is to minimize the $\varrho$-quantile of the end-to-end latency under the energy budget $E_{\max}$. The parameter $\varrho$ can also be viewed in terms of latency violation probability, since the node may continue retransmitting packets until they are successfully received. In this case, the reliability requirement is embedded in the objective, since $F_T^{-1}(\varrho)$ is the latency threshold that is met with probability $\varrho$. Note that, in this case, we do not explicitly model the node's behavior after a latency violation—whether it enters an error state or continues transmitting—as this aspect is beyond the scope of the present work and is left for future studies involving multiple nodes sharing communication resources.
In the time-constrained scenario, the latency budget is fixed in advance, and the objective is to minimize the expected energy while ensuring that compression and transmission are completed within that budget with probability at least $\varrho$. Reliability therefore appears explicitly through the constraint $(1-\epsilon_c)(1-\epsilon_{\mathrm{tx}})\geq\varrho$. Thus, both formulations capture the same end-to-end energy-latency-reliability trade-off, but reliability enters the two problems differently, because one formulation optimizes the latency threshold, whereas the other fixes that threshold and minimizes energy.

\subsection{Energy-Constrained Scenario}\label{Powe_Constrainted_Scenario_Opt}
In this scenario, the sensor is allocated a dedicated \AZ{wireless channel with bandwidth $B$} and transmits without scheduling delays or resource contention. The transmit power is fixed and equal to $P_{\mathrm{tx}}$, as we aim to analyze the general communication and computation trade-off under fixed physical layer parameters. Consequently, the energy allocated per transmitted bit is obtained using \eqref{eq:energy_efficiency_2}. \AZ{We assume a data block is compressed and immediately transmitted. The end-to-end latency, \AM{denoted by $T$}, is then given by the sum of compression, transmission, and decompression times, as illustrated in Fig.~\ref{fig:first_scenario}}. 
\FC{Our objective is to determine} the optimal parameter configurations $\{f_{c},Q\}$ to \AM{minimize the $\varrho$-quantile of the latency \cite{suman2023statistical},} for a given energy budget $E_{\mathrm{max}}$. 

\AZ{Formally, denoting by $F_{T}(t)=\Pr[T\leq t]$ the \gls{cdf} of \AM{ $T$}, the} resulting optimization problem can then be expressed as:
\begin{problemeq}
    \label{eq:Power_Const_Opt}
\begin{split}
    \min_{f_{c},Q} & \ \ \ F^{-1}_T(\varrho), \\
    \mathrm{s.t. } & \ \ \ E(f_{c},Q) \leq E_{\mathrm{max}},
\end{split}
\end{problemeq}
\AM{where the objective function is the $\varrho$-quantile of end-to-end latency $T$. In the following, we will show that \eqref{eq:Power_Const_Opt} can be efficiently solved via a binary search over the optimization variable $Q$.}
The operating point is determined before compression begins by selecting the compression ratio $Q$ and the \gls{cpu} frequency $f_c$. In the slow fading regime, this selection is based on the instantaneous \gls{snr} $\gamma$, which is estimated beforehand and assumed to remain constant throughout. In the fast fading regime, the packet-level channel realizations vary independently and are not known in advance; the operating point is therefore selected from the average \gls{snr} and the fading distribution. The resulting latency quantile is evaluated over the corresponding transmission outcomes.

\begin{figure}[t]
    \centering
    \begin{tikzpicture}
\usetikzlibrary{decorations.pathreplacing}

    \draw[->] (-3,0) -- (3,0) node[pos=1.0, below] {$t$ [s]};
    \draw[->] (-3,0) -- (-3, 1.5) node[pos=0.9, left] {$B$ [Hz]};

    \draw[black, fill=color1] (-2,0) rectangle (-0.7,0.7) node[pos=.5] {$T_c$};
    \draw[black, fill=color2] (-0.7,0) rectangle (1,0.7) node[pos=.5] {$T_{\mathrm{tx}}$};
    \draw[black, fill=color1] (1,0) rectangle (1.6,0.7) node[pos=.5] {$T_d$};
    
    \draw[->] (-2.5,-0.4) -- (-2,0) node[pos=.0, below] {Compression starts};
    \draw[->] (-0.9,-0.8) -- (-0.7,0) node[pos=.0, below] {\ \ \ Transmission starts};
    \draw[->] (0.6,-0.4) -- (1,0) node[pos=.0, below] {Data is received};
    \draw[->] (2.0,-1.2) -- (1.6,0) node[pos=.0, below] {Data is decompressed};

    \draw[->] (-2,1.5) -- (-2,0.35) node[pos=.0, above] {Data requested};
    \draw[->] (1.6,0.35) -- (1.6,1.5) node[pos=1.0, above] {Data delivered};

 \draw [decorate,decoration={brace,amplitude=5pt,raise=-3ex}]
  (-2,1.3) -- (1.6,1.3) node[midway]{End-to-end Latency};
\end{tikzpicture}
    \vspace{-0.2cm}
    \caption{Energy-Constrained scenario.}
    \label{fig:first_scenario}
    \vspace{-0.4cm}
\end{figure}
\subsubsection{Timing}
Let us consider a data block of $D$ bits, whose compression time $T_{c}$ is obtained from equation \eqref{eq:compression_time}. After compression, the $\lceil D/Q \rceil$ bits are segmented into packets of equal length $n_{p}$, with padding applied as needed to obtain an integer number $N$ of packets, with
\begin{equation}\label{eq:number_of_packets}
    N=\left\lceil D(Qn_p)^{-1}\right\rceil\,.
\end{equation}
In both the slow and fast fading cases, packet transmissions are treated as \gls{iid} Bernoulli trials: in the slow fading case, the \gls{snr} is fixed and the outcome of the transmission is determined by the realization of the channel noise, while in the fast fading case, the channel realization for each packet is also \gls{iid}, following the exponential fast fading distribution.
 \FC{The number of attempts necessary to transmit a single packet thus follows a geometric distribution with parameter $1-\varepsilon$\cite{suman2023statistical}. \rev{In the fast-fading case, although the conditional packet error probability varies with the channel realization, each transmission attempt has the same fading-averaged error probability under the assumed i.i.d. fading model. Therefore, the transmission outcomes remain i.i.d. Bernoulli trials and the number of attempts remains geometrically distributed.} Accordingly,} the probability of requiring exactly $K$ attempts to correctly receive a packet is given by:
\begin{equation}\label{eq:rx_attempts}
\Pr[K = k] = \varepsilon^{k-1} (1-\varepsilon),\, k\geq 1,
\end{equation}
Consequently, the time required to successfully transmit one packet, denoted by $\Tau$, is a random variable with the same distribution as $K$, scaled by \AM{the transmission} time $t_p$. The expected value and variance of variable $\Tau$ are expressed as:

\begin{equation}\label{eq:Total_T_Stats}
\E{\Tau} = \frac{t_{p}}{(1-\varepsilon)},\quad
\E{(\Tau- \E{\Tau})^2} = \frac{\varepsilon t_p^2}{(1-\varepsilon)^2}. 
\end{equation}

Following equations \eqref{eq:transmit_time} and \eqref{eq:Total_T_Stats}, the total
channel occupation time required to successfully transmit a data block of $N$
packets is given by
\begin{equation}
T_{\mathrm{tx}} = \sum_{i=1}^{N} \Tau_i,
\end{equation}
where $\Tau_i$ denotes the transmission time required for the $i$-th packet and
the $\Tau_i$ are \gls{iid} random variables following a geometric distribution
scaled by the packet duration $t_p$. Equivalently, the total number of
transmission attempts, $N_{\mathrm{tx}}$, follows a negative binomial distribution with parameters $\{N,\varepsilon\}$, representing the number of attempts required to achieve $N$ successful packet deliveries. Hence, the total
channel occupation time is
\begin{equation}
    T_{\mathrm{tx}} = N_{\mathrm{tx}} \cdot t_p .
\end{equation}
The above expression accounts for the time during which the transmitter is
actively sending packets and is therefore used to compute the transmission
energy consumption. However, we need to account for the use of guard intervals in the fast fading scenario, which ensure independent channel realizations between retransmission attempts. Therefore, the end-to-end latency must also account for the
possible waiting time required to obtain an independent fading realization.

Let $T_{\mathrm{rt}}$ denote the time interval between two consecutive
transmission attempts. In the slow fading case, we simply have $T_{\mathrm{rt}}=t_p$. On the other hand, in the fast fading case we need to ensure $T_{\mathrm{rt}}\geq \max(t_p,T_{\mathrm{coh}})$. The
latency contribution due to retransmissions can then be expressed as
\begin{equation}
    T_{\mathrm{lat,tx}} = N_{\mathrm{tx}} T_{\mathrm{rt}},
\end{equation}

whereas the actual channel occupation time remains $T_{\mathrm{tx}} = N_{\mathrm{tx}}t_p$. The end-to-end latency is given
by
\begin{equation}\label{eq:first_scenario_time}
    T = T_c + N_{\mathrm{tx}}T_{\mathrm{rt}} + T_d = \left(1+\frac{\zeta f_c}{f_b}\right) T_c + T_{\mathrm{lat,tx}} ,
\end{equation}
where we assumed $T_{d} =\frac{\zeta f_{c}}{f_{b}}T_c$,  with \( f_{b} \) denoting the \gls{bs} \gls{cpu} frequency. The end-to-end latency random variable, \( T \), is thus obtained as the sum of two components: \AZ{a scaled version of} \( T_{c} \), which is modeled as a Gamma-distributed random variable, and \( T_{\mathrm{lat,tx}} \), which follows a negative binomial distribution scaled by the duration $T_{\mathrm{rt}}$. As a result, the distribution of \( T \) is not available in closed form and must be characterized numerically or through approximation techniques. To this end, a recent work~\cite{li2025unified} investigated \gls{spa} to obtain the \gls{cdf} of the sum of random variables describing different latency components \FC{in a more general case. However, we can exploit our knowledge of the distributions of the various components to propose} a more tractable upper bound and provide \FC{reliability} guarantees on the end-to-end latency $T$. Following \cite{li2025unified}, we observe that for $N \gg 1$ and sufficiently small $T_{\mathrm{rt}}$, the time $T_{\mathrm{lat,tx}}$ can be approximated by a Gaussian distribution as: 
\begin{equation}
\label{eq:channel_gauss_approx}
\N\left(\mu_{\mathrm{tx}}=\frac{NT_{\mathrm{rt}}}{(1-\varepsilon)}, \sigma_{\mathrm{tx}}^{2}=\frac{NT_{\mathrm{rt}}^{2}\varepsilon}{(1-\varepsilon)^2}\right).
\end{equation} 
As a consequence, the end-to-end latency can be approximated as $T\simeq\left(1+ ({\zeta f_{c}}/{f_{b}}) \right)T_c + T^G_{\textrm{tx}}$, where $T^G_{\textrm{tx}}~\sim~\N(\mu_{\textrm{tx}},\sigma^2_{\textrm{tx}})$. The timing components are shown in Fig.~\ref{fig:first_scenario}.

\subsubsection{Energy}
Similar to the end-to-end latency, the system's overall average energy consumption $E$ is given by the sum of the computation and communication energy:
\begin{equation}
    E=E_c+E_{\mathrm{tx}} = P(f_{c})\E{T_c} + E_{\mathrm{tx}}. 
\end{equation}
Here, we omitted the decompression energy\FC{, as it does not affect the transmitter's battery life: the \gls{bs} is assumed to be connected to a power source, and is thus irrelevant to our study. The compression energy} $E_c$ can be easily computed following the discussion in Section \ref{sec:comp_model}. On the other hand, the communication energy is:
\begin{equation}~\label{eq:comm_energy}
E_{\textrm{tx}}=\frac{n_p\E{T_{\mathrm{tx}}}}{\eta(\varepsilon)t_p},
\end{equation}
as the energy required to transmit a single packet is ${n_p}/{\eta(\varepsilon)}$. 

\subsubsection{Optimization}
In the following, we find an approximate solution to problem~\eqref{eq:Power_Const_Opt}. 
We first express the energy constraint \AM{in problem \eqref{eq:Power_Const_Opt}} as an explicit function of $f_c$ and $Q$:
\begin{equation}\label{eq:constraint}
    \frac{(e^{\psi Q } - e^\psi)D P_{s,\mathrm{max}} f_c^2}{f_\mathrm{max}^3} + \frac{D}{(1-\varepsilon)Q \eta(\varepsilon)} \leq E_{\mathrm{max}}.
\end{equation}
 Then, we express the average number of transmitted packets, including \AM{retransmissions}, as $N/(Q(1-\varepsilon))$, omitting the ceiling operation for better tractability. 
The following lemma can then be obtained by analyzing the behavior of the utility function and the constraint in \eqref{eq:constraint} with respect to $f_c$. 
\begin{lemma}
\label{lem:eq_const}
    For any optimal pair $(f_c^*,Q^*)$, the constraint in \eqref{eq:constraint} is always satisfied with equality leading to the following relationship between $f_c^*$ and $Q$
    \begin{equation}\label{eq:f_c_star}
        f_c^* = \sqrt{\frac{E_{\mathrm{max}} - \frac{D}{(1-\varepsilon)Q\eta(\varepsilon)}}{(e^{\psi Q } - e^\psi) D P_{s,\mathrm{max}} / f^3_{\mathrm{max}}}}\,.
    \end{equation}
    \end{lemma} 
    \begin{IEEEproof}
    As the computation frequency $f_c$ increases, the computation time monotonically decreases, and thus the overall latency decreases. \FC{Since the computation time follows a Gamma distribution, scaled by the inverse of the frequency, we can easily prove that $\forall f_{c,1}> f_{c,2} \ \mathrm{and} \ \forall  \varrho$, it holds $F^{-1}_{T_c}(\varrho, f_{c,1})~<~F^{-1}_{T_c}(\varrho, f_{c,2}) $,  i.e., the computation time using $f_{c,2}$ has} first order stochastic dominance with respect to $f_{c,1}$. \FC{By definition, any quantile $\varrho$ of the computation latency distribution is then a monotonically decreasing function of $f_c$, and the condition is extended to strict monotonicity if $Q>1$. As the communication latency for a fixed $Q$ is independent of $f_c$, this also holds for the overall latency.} 
    On the other hand, the energy increases monotonically with $f_c$. Choosing the highest possible frequency, i.e., satisfying the constraint at equality, is thus optimal for any set value of $Q^*$. \AZ{Setting  \eqref{eq:constraint} as an equality and solving for $f_c$, we obtain \eqref{eq:f_c_star}.}
\end{IEEEproof}
Using \eqref{eq:f_c_star},  problem~\eqref{eq:Power_Const_Opt} can be rewritten as an \FC{ optimization problem with a single variable, $Q$}. 
\AM{However, the end-to-end latency remains intractable, as it is the sum of a Gamma-distributed and a Gaussian-distributed random variable, both of which are functions of the compression parameter \( Q \). We then solve the problem for a tractable upper bound for the \( \varrho \)-quantile of the latency distribution, denoted by \( F_T^{-1}(\varrho) \).}

\PT{As both the Gaussian and Gamma random variables are unimodal and log-concave, their quantiles have the sub-additivity property~\cite{bagnoli2006log}, and we can write
    \begin{equation}
    \label{eq:quantile_bound}
        F_T^{-1}(\varrho) \leq F_{T_c}^{-1}(\varrho) + F_{T_{\mathrm{tx}}^G}^{-1}(\varrho).
    \end{equation}}
\PT{We then} relax~\eqref{eq:Power_Const_Opt} to the minimization of \PT{the} upper bound of the latency with respect to $Q$. \PT{We will study the tightness of the quantile bound under practical conditions in the following.} The \AM{relaxed} problem \AM{is formulated as:}
\begin{problemeq}\label{eq:unconstrained_prob}
    \min_{Q} F_{T_c}^{-1}(\varrho,Q,f_c^*(Q)) + F_{T_{\mathrm{lat,tx}}}^{-1}(\varrho,Q,f_c^*(Q)),
\end{problemeq}
where  \AM{Lemma~\ref{lem:eq_const} is used to express the optimal compression frequency \( f_c^* \) as a function of \( Q \). Subsequently, we investigate the convexity of problem \eqref{eq:unconstrained_prob} with respect to $Q$.}
\begin{theorem}
\label{th:convex}
    Problem \eqref{eq:unconstrained_prob} is convex with respect to $Q$, under the condition 
    \begin{equation}
    \label{eq:cond}
    Q^23\psi^2e^\psi -2QE_{\mathrm{max}} + \frac{D}{(1-\varepsilon)\eta(\varepsilon)} \geq 0.
    \end{equation}
    \proof To prove that the unconstrained problem is convex, \AM{ it suffices to show} that both quantiles are convex with respect to $Q$. \FC{The statement of the theorem then trivially follows,} given that the sum of two convex functions is itself convex. The condition in~\eqref{eq:cond} ensures the convexity of the quantile \AM{function associated with the Gamma-distributed compression time; a detailed analytical derivation of this condition is provided in Appendix~\ref{appendix:convexity}}. On the other hand, the quantile \AM{function of the Gaussian-distributed approximate \PT{communication} time }can be written as $F_{T^G_{\textrm{tx}}}^{-1}(\varrho)=\mu_{\mathrm{tx}}+z(\varrho)\sigma_{\mathrm{tx}}$, where $z(\varrho)$ is the probit function, \AM{i.e., the inverse \gls{cdf} of the standard normal distribution}. 
Substituting the expressions for \( \mu_{\mathrm{tx}} \) and \( \sigma_{\mathrm{tx}} \), we obtain:
\begin{equation}\label{eq:tcomp_quantile}
    F_{T^G_{\textrm{tx}}}^{-1}(\varrho) = \frac{N T_{\mathrm{rt}}}{1 - \varepsilon} + z(\varrho) \cdot T_{\mathrm{rt}} \cdot \sqrt{\frac{\varepsilon N}{(1 - \varepsilon)^2}}.
\end{equation}
Since \( N = \frac{D}{Q n_p} \) is a non-increasing and convex function of \( Q \) for \( Q > 0 \), and both the mean and standard deviation of the Gaussian distribution are affine in \( N \), it follows that \( F_{T^G_{\textrm{tx}}}^{-1}(\varrho) \) is convex in \( Q \):
\begin{equation}
\label{eq:tx_quantile}
    F_{T^G_{\textrm{tx}}}^{-1}(\varrho) = z(\varrho)\sqrt{\frac{DT_{\mathrm{rt}}^2 \varepsilon}{Qn_p(1-\varepsilon)^2}} + \frac{DT_{\mathrm{rt}}}{Qn_p(1-\varepsilon)}.
\end{equation}
\AZ{Since  \( F_{T^G_{\textrm{tx}}}^{-1}(\varrho) \) is the sum of two functions, each of which is convex in \( Q \) for \( Q > 0 \), it is itself a convex function of $Q$, which concludes the proof.  }
\end{theorem}
Although \FC{we proved that the problem is convex under reasonable conditions}, \AZ{we could not derive a closed-form expression for the minimum latency}. However, \AM{the optimization problem in the energy-constrained scenario can be efficiently solved using a binary search algorithm, with computational complexity $
O\left(\log\left(\frac{Q_{\mathrm{max}} - 1}{\theta}\right)\right),$
where
\( \theta \) is the desired precision of the solution. Alternatively, the problem can be tackled using Lagrangian methods or other standard convex optimization solvers.}

\begin{figure}[!t]
    \centering
    \begin{tikzpicture}
    \draw[->] (-3,0.5) -- (3,0.5) node[pos=1.0, below] {$t$ [s]};
    \draw[->] (-3,0.5) -- (-3, 1.5) node[pos=0.5, left] {Computation};

    \draw[->] (-3,2) -- (3,2) node[pos=1.0, below] {$t$ [s]};
    \draw[->] (-3,2) -- (-3, 3) node[pos=0.5, left] {$B$ [Hz]};

    \draw[black, fill = lightgray] (-2.5,2) rectangle (-0.7,2.7) node[pos=.5] {TS 1};
    \draw[black, fill = color2] (-0.7,2) rectangle (1,2.7) node[pos=.5] {TS 2};
    \draw[black, fill= lightgray] (1,2) rectangle (2.7,2.7) node[pos=.5] {TS 3};
    
    \draw[->] (-2.5,0.1) -- (-2,0.5) node[pos=.0, below] {Compression starts};
    \draw[->] (-0.9,-0.3) -- (-0.7,0.5) node[pos=.0, below] {\ \ \ Transmission starts};
    \draw[->] (0.6,0.1) -- (1,0.5) node[pos=.0, below] {Data is received};
    \draw[->] (2.0,-0.7) -- (1.6,0.5) node[pos=.0, below] {Data is decompressed};

    \draw[black, fill = color1] (-2,0.5) rectangle (-0.9,1) node[pos=.5] {$T_c$};
    \draw[black, fill = color1] (1,0.5) rectangle (1.6,1) node[pos=.5] {$T_d$};

    \draw[dashed] (-0.7, 3) -- (-0.7, 0);
    \draw[dashed] (1.0, 3) -- (1.0, 0);
    \draw[dashed] (-2.0, 3) -- (-2.0, 0);
    \draw[dashed] (1.6, 3) -- (1.6, 0);

    \draw[decorate,decoration={brace,amplitude=5pt,raise=-3ex}]
  (-2,1.6) -- (-0.7,1.6) node[midway]{$T_{\mathrm{comp}}$};
  \draw[decorate,decoration={brace,amplitude=5pt,raise=-1.5ex}]
  (1,1.6) -- (-0.7,1.6) node[pos=.5, below]{\begin{tabular}{c} $T_{\mathrm{slot}}$ \\ $(1-\alpha)T$ \end{tabular}};

   \draw [decorate,decoration={brace,amplitude=5pt,raise=-3ex}]
  (-2,3.3) -- (1.,3.3) node[midway]{$T$};

\end{tikzpicture}
\vspace{-1.1cm}
    \caption{Time-Constrained scenario.}
    \label{fig:scheme_second_scenario}
    \vspace{-0.4cm}
\end{figure}

\subsection{Time-Constrained Scenario}\label{sec:scen2}
In this second target scenario, we change the access model by dividing the channel into time-frequency slots. The objective is to design a system \AM{that ensures the total time for compression and transmission does not exceed a fixed maximum duration $T$ in order to meet \gls{qaoi} requirements}. \PT{Note that we are not explicitly modeling the reception and processing of the query. This corresponds to setting (1) a specific time instant by which the transmission should be finalized; and (2) a maximum value of the age of the information delivered at that instant, which is the \gls{qaoi}}. Compression and communication must each occur within pre-allocated time slots, as the diagram in Fig.~\ref{fig:scheme_second_scenario} shows. \AZ{This model could be used in scenarios  where multiple sensors are assigned some communication resources to transmit their data, so that all transmissions are scheduled and avoid any mutual interference.} \rev{In the time-constrained scenario, slow and fast fading are treated as in Sec.~\ref{sec:comm_model}, using respectively the fixed-SNR and fading-averaged packet error probabilities, together with the corresponding inter-transmission interval.}

\par \AZ{The \gls{qaoi}  here spans across three time slots: the transmission slot, used to send the compressed data, and the slot before and after that, which are used to compress and decompress data before and after transmission, respectively. An illustrative example is shown in Fig. \ref{fig:scheme_second_scenario}, \PT{where it is assumed that the data is sampled just at the instant when the compression starts}}. \AZ{While in the energy-constrained scenario the energy budget, which was our constraint, could be optimally divided between the compression and communication operations, here the time constraint $T$ must necessarily fall within the compression and communication slots, which are predefined. However, within these boundaries, we can still adjust the compression  time to maximize the overall \gls{ee}. } 
\AZ{We then introduce an additional} optimization variable $\alpha \in [0,1)$, such that $T_{\textrm{comp}}=\alpha T$ is the fraction of time allocated for compression, while $T_{\mathrm{comm}}=(1-\alpha)T$ is the communication slot. Note that we consider only the compression time in the optimization, as the decompression time is negligible in comparison. Furthermore, due to the computing capabilities at the \gls{bs}, decompression is assumed to be completed within the allocated slot whenever compression and transmission succeed. Hence, although $T$ is not strictly equal to the \gls{qaoi}, it is equivalent in this context. The probabilities that compression and transmission fail to finish within the slot are denoted by $\varepsilon_c$ and $\varepsilon_{\textrm{tx}}$, respectively.
Considering that these operations are independent, the overall reliability of the communication can then be expressed as the product of the success probabilities of the compression and transmission steps, that is $P_{\rm succ}=(1-\varepsilon_c)(1-\varepsilon_{\textrm{tx}})$.
The optimization goal is then to minimize the average energy consumption while ensuring that the system reliability
is at least $\varrho$:
\begin{problemeq}
\label{eq:Time_Const_opt}
\begin{split}
        \min_{f_{c},Q, \alpha} 
        &E_c(f_{c},Q)+E_{\mathrm{tx}}(P_{\mathrm{tx}},Q),\\
        \mathrm{s.t. \ }&\AM{F_{T_{c}}\left(T_{\textrm{comp}}\right)\geq 1-\varepsilon_c,} \\
        &\AM{F_{T_{\textrm{tx}}^G}\left(T_{\textrm{comm}}\right)\geq 1-\varepsilon_{\textrm{tx}}, }\\
        &(1-\varepsilon_c)(1-\varepsilon_{\AM{\textrm{tx}}}) {\geq} \varrho\,. 
\end{split}
\end{problemeq}
\subsubsection{Timing}
As shown in Fig.~\ref{fig:scheme_second_scenario}, the compression \AM{operation} must terminate before the transmission begins, \AM{i.e., $T_{c}<T_{\textrm{comp}}$}. This sets the first constraint of the system, requiring data to be reliably compressed before the start of the assigned time-frequency slot. Fixing $Q$ and $f_{c}$ implicitly leads to a fixed contribution to the \gls{qaoi}, provided that the operation concludes in time. The overall time between the beginning of compression at the transmitter and the decoding of the data at the receiver is the  \gls{qaoi}~\cite{Federico@QAoI}, as it represents the freshness of the information available to the receiver if the time slot allocation aligns with client requests. \PT{The \gls{qaoi} is the result of the scheduling choices of compression and communication to meet some requests on the age of the delivered data and the reliability constraint}. The \gls{qaoi} is hence expressed as:
\begin{equation}\label{eq:qaoi}
    \mathrm{\gls{qaoi}} =T_{\textrm{comp}} + T_{\mathrm{\AM{comm}}} + T_{d,\mathrm{max}},
\end{equation}
where $T_{d,\mathrm{max}}$ is the time allocated for decompression, which, as described in Sec.~\ref{sec:comp_model}, is set as a fraction of the compression-time budget. Since the \gls{bs} \gls{cpu} frequency satisfies $f_b \gg f_c$ and $\zeta \ll 1$, $T_{d}$ is non-binding in our regime and is omitted from the optimization objective.
Since the sampling and transmission instants are \AM{predetermined by virtue of allocation of the respective slots}, compression must happen within $T_{\textrm{comp}}$, or the sensor will be unable to transmit in the transmission slot $T_{\textrm{comm}}$.  In this case, the transmitter receives a feedback on the success of \AM{transmitted packets} only after the slot terminates and, consequently, it cannot rely on retransmissions. As such, we propose packet-level coding to reduce the message error probability. Efficient packet-level codes enable decoding of a block of $N$ original packets if at least $N$ out of the $N_{\mathrm{tx}} > N$ transmitted packets are correctly received~\cite{rizzo1997effective}. \AZ{Assuming a Bernoulli error process,} the \AM{transmission error probability is obtained as}:
\begin{equation}\label{eq:error_message_prob}
    \varepsilon_{\mathrm{tx}} =  \textstyle\sum_{h=0}^{N-1} \binom{N_{\mathrm{tx}}}{h} (1-\varepsilon)^h \varepsilon^{N_{\mathrm{tx}}-h}\,,
\end{equation}
\AZ{where $\varepsilon$ is the channel packet error probability.}

\begin{figure*}
\centering
    \subfloat[Approximation error with to $\varrho$. \label{fig:err_rel}]{
\begin{tikzpicture}

\definecolor{darkgray176}{RGB}{176,176,176}

\definecolor{crimson2143940}{RGB}{240, 249, 33}
\definecolor{darkorange25512714}{RGB}{253,180,47}
\definecolor{forestgreen4416044}{RGB}{237,121,83}
\definecolor{mediumpurple148103189}{RGB}{204,71,120}
\definecolor{orchid227119194}{RGB}{156, 23, 158}
\definecolor{sienna1408675}{RGB}{92,1,166}
\definecolor{steelblue31119180}{RGB}{13,8,135}

\begin{axis}[
ybar,
width=25em,
height=10em,
grid=both,
grid style={line width=.1pt, draw=gray!10},
major grid style={line width=.2pt,draw=gray!40},
minor tick num=1,
xlabel={$\varrho$},
ylabel={Approx. error (\%)},
x grid style={darkgray176},
xmin=0.5, xmax=5.5,
xtick style={color=black},
y grid style={darkgray176},
ymin=0.0, ymax=1.5,
ytick style={color=black},
xtick={1, 2, 3, 4, 5},
xticklabels={0.9, 0.99, 0.999, 0.9999, 0.99999},
legend style={draw=white!15!black, font=\scriptsize, at={(0.21, 0.98)}, anchor=north east, /tikz/every even column/.append style={column sep=0.2em}},
legend columns=1
]
\addplot [fill=color7]
table {figures/error_reliability.dat};
\addplot [fill=color2]
table {figures/error_monte_carlo_rho.dat};
\legend{SPA, MC}
\end{axis}
\end{tikzpicture}}
    \subfloat[Approximation error with respect to $\varepsilon$. \label{fig:err_eps}]{
\begin{tikzpicture}

\definecolor{darkgray176}{RGB}{176,176,176}

\definecolor{crimson2143940}{RGB}{240, 249, 33}
\definecolor{darkorange25512714}{RGB}{253,180,47}
\definecolor{forestgreen4416044}{RGB}{237,121,83}
\definecolor{mediumpurple148103189}{RGB}{204,71,120}
\definecolor{orchid227119194}{RGB}{156, 23, 158}
\definecolor{sienna1408675}{RGB}{92,1,166}
\definecolor{steelblue31119180}{RGB}{13,8,135}

\begin{axis}[
ybar,
width=25em,
height=10em,
grid=both,
grid style={line width=.1pt, draw=gray!10},
major grid style={line width=.2pt,draw=gray!40},
minor tick num=1,
xlabel={$\varepsilon$},
ylabel={Approx. error (\%)},
x grid style={darkgray176},
xmin=0.5, xmax=3.5,
xtick style={color=black},
y grid style={darkgray176},
ymin=0.0, ymax=3.0,
ytick style={color=black},
xtick={1, 2, 3},
xticklabels={0.1, 0.01, 0.001},
legend style={draw=white!15!black, font=\scriptsize, at={(0.99, 0.98)}, anchor=north east, /tikz/every even column/.append style={column sep=0.2em}},
legend columns=1
]
\addplot [fill=color7]
table {figures/error_epsilon.dat};
\addplot [fill=color2]
table {figures/error_monte_carlo_epsilon.dat};
\legend{SPA, MC}
\end{axis}
\end{tikzpicture}}
    \caption{Approximation error on the channel latency introduced by \eqref{eq:channel_gauss_approx} with respect to the reliability $\varrho$ and to the error probability $\varepsilon$.}
\end{figure*}

\AM{Building on this, the resulting \gls{qaoi} is a constant as problem~\eqref{eq:Time_Const_opt} is optimized to restrict both compression and communication with a predetermined time interval.}
\subsubsection{Energy}
\FC{Sec.~\ref{sec:comp_model} models the compression time as a Gamma distribution. In this scenario, however, we assume that } the compression procedure \AZ{is forced to terminate, even if incomplete, at the beginning of the transmission slot.} \FC{This results in a \textit{truncated} Gamma distribution of the compression time.}
The energy for compression $E_{c}$ is then computed as
\begin{equation}
    E_c = \big((1-\varepsilon_c)\E{T_c \mid T_c <  T_{\textrm{comp}}} + \varepsilon_c  T_{\textrm{comp}} \big) P_c(f_c).
\end{equation}
 The \FC{expected value of the truncated Gamma distribution admits a closed-form solution~\cite{alai2013lifetime}.} In the following, we derive this solution with respect to the parameters considered here as:
\begin{align}
    \E{T_c \lvert T_c < T_{\textrm{comp}}} = \frac{\beta D}{f_c} \frac{\gamma(\kappa + 1, \frac{\beta D}{f_c T_{\textrm{comp}}})}{\gamma(\kappa , \frac{\beta D}{f_c T_{\textrm{comp}}})} = \bar{T_c}^{\mathrm{trunc}},
\end{align}
where $\gamma(s,x) = \int_0^x t^{s-1}e^{-t} dt$ is the lower incomplete gamma function and $\beta = (e^{\psi Q} - e^{\psi})/\kappa $. It follows that the average energy consumption for compression is given by $E_{c}(f_c,Q) = P(f_c) \bar{T_c}^{\mathrm{trunc}}$. On the other hand, since retransmissions are not employed, the number of transmitted packets is fixed as: 
\begin{equation}
\label{eq:n_tx}
    N_{\mathrm{tx}}= \left\lfloor T_{\mathrm{rt}}^{-1}(1-\alpha)T \right\rfloor.
\end{equation} 
\FC{Therefore, $T_{\textrm{comm}}$ is fully used to maximize reliability.} The communication energy is then determined by~\eqref{eq:comm_energy}. 
\subsubsection{Optimization}

Similar to~\eqref{eq:Power_Const_Opt}, problem~\eqref{eq:Time_Const_opt} can be solved for a given {\gls{qaoi}}, with $f_c$, $Q$ and $\alpha$ as optimization variables. In this case we can note that, for any pair $(Q,\alpha)$, we can always decrease the processing frequency $f_c$ until the constraint on overall reliability is strict, i.e.,  satisfied with equality. The proof is complementary to Lemma~\ref{lem:eq_const}, as both the compression energy $E_c$ and the probability of successful compression $(1-\varepsilon_c)$ decrease monotonically as $f_c$ increases.

\begin{figure*}
\centering
    \subfloat[Reliability $\varrho$. \label{fig:q_rho}]{
\begin{tikzpicture}

\definecolor{darkgray176}{RGB}{176,176,176}

\definecolor{crimson2143940}{RGB}{240, 249, 33}
\definecolor{darkorange25512714}{RGB}{253,180,47}
\definecolor{forestgreen4416044}{RGB}{237,121,83}
\definecolor{mediumpurple148103189}{RGB}{204,71,120}
\definecolor{orchid227119194}{RGB}{156, 23, 158}
\definecolor{sienna1408675}{RGB}{92,1,166}
\definecolor{steelblue31119180}{RGB}{13,8,135}

\begin{axis}[
ybar,
width=14em,
height=10em,
grid=both,
grid style={line width=.1pt, draw=gray!10},
major grid style={line width=.2pt,draw=gray!40},
minor tick num=1,
xlabel={$1-\varrho$},
ylabel={Bound gap (\%)},
x grid style={darkgray176},
xmin=0.5, xmax=5.9,
xtick style={color=black},
y grid style={darkgray176},
ymin=0.0, ymax=2.5,
ytick style={color=black},
xtick={1, 2, 3, 4.1, 5.4},
xticklabels={$10^{-1}$,$10^{-2}$,$10^{-3}$,$10^{-4}$,$10^{-5}$},
legend style={draw=white!15!black, font=\scriptsize, at={(0.02, 0.98)}, anchor=north west, /tikz/every even column/.append style={column sep=0.2em}},
legend columns=1
]
\addplot [fill=color2]
coordinates {
(1, 0.54709735 ) 
(2, 1.21430005 )
(3, 1.69393464 )
(4.1, 2.07054827 )
(5.4, 2.4889212)
};
\legend{MC}
\end{axis}
\end{tikzpicture}}
    \hspace{-0.2cm}
    \subfloat[Packet erasure $\varepsilon$. \label{fig:q_eps}]{
\begin{tikzpicture}

\definecolor{darkgray176}{RGB}{176,176,176}

\definecolor{crimson2143940}{RGB}{240, 249, 33}
\definecolor{darkorange25512714}{RGB}{253,180,47}
\definecolor{forestgreen4416044}{RGB}{237,121,83}
\definecolor{mediumpurple148103189}{RGB}{204,71,120}
\definecolor{orchid227119194}{RGB}{156, 23, 158}
\definecolor{sienna1408675}{RGB}{92,1,166}
\definecolor{steelblue31119180}{RGB}{13,8,135}

\begin{axis}[
ybar,
width=14em,
height=10em,
grid=both,
grid style={line width=.1pt, draw=gray!10},
major grid style={line width=.2pt,draw=gray!40},
minor tick num=1,
xlabel={$\textcolor{white}{Q}\varepsilon\textcolor{white}{Q}$},
x grid style={darkgray176},
xmin=0.5, xmax=3.5,
xtick style={color=black},
y grid style={darkgray176},
ymin=0.0, ymax=1.5,
ytick style={color=black},
xtick={1, 2, 3},
xticklabels={0.1, 0.01, 0.001},
legend style={draw=white!15!black, font=\scriptsize, at={(0.99, 0.98)}, anchor=north east, /tikz/every even column/.append style={column sep=0.2em}},
legend columns=1
]
\addplot [fill=color2]
coordinates {
(1, 1.1905)
(2, 0.2106)
(3, 0.0242)
};

\legend{MC}
\end{axis}
\end{tikzpicture}}
    \hspace{-0.2cm}
    \subfloat[Compression ratio $Q$. \label{fig:q_q}]{
\begin{tikzpicture}

\definecolor{darkgray176}{RGB}{176,176,176}

\definecolor{crimson2143940}{RGB}{240, 249, 33}
\definecolor{darkorange25512714}{RGB}{253,180,47}
\definecolor{forestgreen4416044}{RGB}{237,121,83}
\definecolor{mediumpurple148103189}{RGB}{204,71,120}
\definecolor{orchid227119194}{RGB}{156, 23, 158}
\definecolor{sienna1408675}{RGB}{92,1,166}
\definecolor{steelblue31119180}{RGB}{13,8,135}

\begin{axis}[
ybar,
width=14em,
height=10em,
grid=both,
grid style={line width=.1pt, draw=gray!10},
major grid style={line width=.2pt,draw=gray!40},
minor tick num=1,
xlabel={$Q$},
x grid style={darkgray176},
xmin=0.5, xmax=5.5,
xtick style={color=black},
y grid style={darkgray176},
ymin=0.0, ymax=1.5,
ytick style={color=black},
xtick={1, 2, 3, 4, 5},
xticklabels={1.1, 1.2, 1.3, 1.4, 1.5},
legend style={draw=white!15!black, font=\scriptsize, at={(0.99, 0.98)}, anchor=north east, /tikz/every even column/.append style={column sep=0.2em}},
legend columns=1
]
\addplot [fill=color2]
coordinates {
(1, 1.2143404) 
(2, 1.15316125)
(3, 0.81213869)
(4, 0.68442753)
(5, 0.38172698)
};
\legend{MC}
\end{axis}
\end{tikzpicture}}
    \hspace{-0.2cm}
    \subfloat[Computing frequency $f_c$. \label{fig:q_fc}]{
\begin{tikzpicture}

\definecolor{darkgray176}{RGB}{176,176,176}

\definecolor{crimson2143940}{RGB}{240, 249, 33}
\definecolor{darkorange25512714}{RGB}{253,180,47}
\definecolor{forestgreen4416044}{RGB}{237,121,83}
\definecolor{mediumpurple148103189}{RGB}{204,71,120}
\definecolor{orchid227119194}{RGB}{156, 23, 158}
\definecolor{sienna1408675}{RGB}{92,1,166}
\definecolor{steelblue31119180}{RGB}{13,8,135}

\begin{axis}[
ybar,
width=14em,
height=10em,
grid=both,
grid style={line width=.1pt, draw=gray!10},
major grid style={line width=.2pt,draw=gray!40},
minor tick num=1,
xlabel={$f_c$[GHz]},
x grid style={darkgray176},
xmin=0.5, xmax=5.5,
xtick style={color=black},
y grid style={darkgray176},
ymin=0.0, ymax=1.5,
ytick style={color=black},
xtick={1, 2, 3, 4, 5},
xticklabels={0.8, 1, 1.5, 2, 2.5},
legend style={draw=white!15!black, font=\scriptsize, at={(0.99, 0.98)}, anchor=north east, /tikz/every even column/.append style={column sep=0.2em}},
legend columns=1
]
\addplot [fill=color2]
coordinates {
(1, 0.51753872 ) 
(2, 0.52004781 )
(3, 0.75966757 )
(4, 0.83677922 )
(5, 0.97204179)
};
\legend{MC}
\end{axis}
\end{tikzpicture}}
    \caption{Gap on latency introduced by \eqref{eq:quantile_bound} with respect to different system parameters.}
    \label{fig:quantile_bound}
    \vspace{-0.2cm}
\end{figure*}

On the other hand, the utility function and the constraint in \eqref{eq:Time_Const_opt} are not monotonic in $Q$ and $\alpha$, preventing further simplification of the optimization procedure. The optimization problem is not convex and is solved through exhaustive search on the parameter space. However, given a pair $(\alpha,Q)$, it is still possible to perform binary search to find the optimal frequency $f_c^*$, reducing the complexity \FC{of exhaustive search} to:
\begin{equation}
    O\left(\frac{Q_{\mathrm{max}}-1}{\theta^2}\log_2\left(\frac{f_{c,\mathrm{max}} - f_{c,\mathrm{min}}}{\theta}\right)\right).
\end{equation}

It is possible that some values of the parameters does not satisfy the constraints. There are implicit constraints that are not highlighted in problem \eqref{eq:Time_Const_opt} but are useful to understand the optimization problem. For example, we can see that from \eqref{eq:n_tx} and the upper bound on the compression ratio $Q_{\mathrm{max}}$, that there is a maximum value for $\alpha$. \FC{If we compress with the highest possible compression ratio, we obtain the minimum possible number of packets $N^*$: these packets must be transmitted, as transmitting fewer packets would lead to a failure in all cases.}
Specifically, $\alpha_{\mathrm{max}}$ is obtained as:
\begin{equation}
\label{eq:alpha_max}
    \alpha_{\mathrm{max}} = 1- \left \lceil D(Q_{\mathrm{max}}n_p)^{-1} \right \rceil T^{-1}T_{\mathrm{rt}}.
\end{equation}
This allows to set the limits on the parameter space. The optimization procedure searches parameters combinations within these limits. The non-convexity \FC{of this problem} and the relationship between the three optimization parameters are analyzed in the following section.

\section{Results}

\begin{table}[!t]
    \centering
    \caption{Parameters used in the simulation.}
    \begin{tabular}{c|c|c|c}
        Parameter & Value & Parameter & Value \\
        \hline
        $D$ & $500$ [kB] & $d$ & 500 [m] \\
        $\kappa$ & 1.25 & $\ell$ & 3 \\  
        $\lambda $ & $10^{-15}$ [J/bit] & $N_0$ & -110 [dBm/Hz]\\  
        $\nu$ & $10^{-14}$ [J] & $\varepsilon $ & 0.01 \\
        $\psi$ & 3.5 & $K_0$ & -27 [dB]\\
        $P_{\mathrm{tx}}$ & 0.25 [W] & $f_{c}$ & [$0.8$, $2.5$] [GHz] \\     
        $B$ & $20$ [MHz] & $\zeta$ & 0.05 \\
        $P_{s,\mathrm{max}}$ & $5$ [W] & $N_b$ & 30 \\
        $T_{\mathrm{coh}}$ & 30[ms] & $f_D$ & 70[Hz] \\
    \end{tabular}
    \label{tab:parameters}
    \vspace{-0.4cm}
\end{table}

\begin{figure*}
    \label{fig:bound_validation}
    \centering
    \subfloat[$\varrho=0.99$. \label{fig:quantile_bound_99}]{
\usetikzlibrary{fillbetween}
\begin{tikzpicture}

\definecolor{darkgray176}{RGB}{176,176,176}

\definecolor{crimson2143940}{RGB}{240, 249, 33}
\definecolor{darkorange25512714}{RGB}{253,180,47}
\definecolor{forestgreen4416044}{RGB}{237,121,83}
\definecolor{mediumpurple148103189}{RGB}{204,71,120}
\definecolor{orchid227119194}{RGB}{156, 23, 158}
\definecolor{sienna1408675}{RGB}{92,1,166}
\definecolor{steelblue31119180}{RGB}{13,8,135}

\begin{axis}[
width=24em,
height=13em,
grid=both,
grid style={line width=.1pt, draw=gray!10},
minor tick num=1,
xlabel={$Q$},
ylabel={$F_{T}^{-1}(\varrho)$ [s]},
x grid style={darkgray176},
xmin=1.1, xmax=1.35,
xtick style={color=black},
y grid style={darkgray176},
ymin=0.27, ymax=0.32,
ytick style={color=black},
yticklabel style={
        /pgf/number format/fixed,
        /pgf/number format/precision=3
    },
legend style={draw=white!15!black, font=\scriptsize, at={(0.5, 0.99)}, anchor=north, /tikz/every even column/.append style={column sep=0.2em}},
legend columns=1,
set layers,
]
\addplot [thick, name path=upper, color2] table {power_plots/upper_quantile_new.dat};
\addplot [thick, color7] table {power_plots/quantile_mc_new.dat};
\addplot [thick, name path=lower, color3] table {power_plots/lower_quantile_new.dat};

\addplot[
        fill=color2,
        draw=none,
        fill opacity=0.2
    ] fill between[
        of=upper and lower
    ];
\draw[<->, black] (axis cs:1.19,0.2815) -- node[above=5pt, font=\footnotesize] {1\% interval} (axis cs:1.19,0.2785);

\legend{Upper Bound, Monte Carlo, 99\% Upper Bound};
\end{axis}
\end{tikzpicture}}
    \subfloat[$\varrho=0.99999$. \label{fig:quantile_bound_99999}]{
\usetikzlibrary{fillbetween}
\begin{tikzpicture}

\definecolor{darkgray176}{RGB}{176,176,176}

\definecolor{crimson2143940}{RGB}{240, 249, 33}
\definecolor{darkorange25512714}{RGB}{253,180,47}
\definecolor{forestgreen4416044}{RGB}{237,121,83}
\definecolor{mediumpurple148103189}{RGB}{204,71,120}
\definecolor{orchid227119194}{RGB}{156, 23, 158}
\definecolor{sienna1408675}{RGB}{92,1,166}
\definecolor{steelblue31119180}{RGB}{13,8,135}

\begin{axis}[
width=24em,
height=13em,
grid=both,
grid style={line width=.1pt, draw=gray!10},
minor tick num=1,
xlabel={$Q$},
ylabel={$F_{T}^{-1}(\varrho)$ [s]},
x grid style={darkgray176},
xmin=1.0, xmax=1.3,
xtick style={color=black},
y grid style={darkgray176},
ymin=0.29, ymax=0.34,
ytick style={color=black},
yticklabel style={
        /pgf/number format/fixed,
        /pgf/number format/precision=3
    },
legend style={draw=white!15!black, font=\scriptsize, at={(0.5, 0.99)}, anchor=north, /tikz/every even column/.append style={column sep=0.2em}},
legend columns=1,
set layers,
]
\addplot [thick, name path=upper, color2] table {power_plots/upper_quantile_new_99999.dat};
\addplot [thick, color7] table {power_plots/quantile_mc_new_99999.dat};
\addplot [thick, name path=lower, color3] table {power_plots/lower_quantile_new_99999.dat};

\addplot[
        fill=color2,
        draw=none,
        fill opacity=0.2
    ] fill between[
        of=upper and lower
    ];
\draw[<->, black] (axis cs:1.1,0.2977) -- node[above=5pt, font=\footnotesize] {1\% interval} (axis cs:1.1,0.294);

\legend{Upper Bound, Monte Carlo, 99\% Upper Bound};
\end{axis}
\end{tikzpicture}}
    \caption{Visualization of the gap between upper bound in \eqref{eq:quantile_bound} and Monte Carlo estimation of the end-to-end latency.}\vspace{-0.4cm}
    
\end{figure*}

\begin{figure*}
    \centering
    \subfloat[Optimal frequency $f_c^*(Q)$. \label{fig:optim_f}]{
\begin{tikzpicture}

\definecolor{darkgray176}{RGB}{176,176,176}

\definecolor{crimson2143940}{RGB}{240, 249, 33}
\definecolor{darkorange25512714}{RGB}{253,180,47}
\definecolor{forestgreen4416044}{RGB}{237,121,83}
\definecolor{mediumpurple148103189}{RGB}{204,71,120}
\definecolor{orchid227119194}{RGB}{156, 23, 158}
\definecolor{sienna1408675}{RGB}{92,1,166}
\definecolor{steelblue31119180}{RGB}{13,8,135}

\begin{axis}[
width=24em,
height=13em,
grid=both,
grid style={line width=.1pt, draw=gray!10},
major grid style={line width=.2pt,draw=gray!40},
minor tick num=1,
xlabel={$Q$},
ylabel={$f_c^*$ [GHz]},
x grid style={darkgray176},
xmin=1.0, xmax=1.5,
xtick style={color=black},
y grid style={darkgray176},
ymin=0.5, ymax=2.5,
ytick style={color=black},
legend style={at={(0.5,1.15)}, anchor=north,legend columns=2},
legend style={draw=white!15!black, font=\scriptsize, at={(1.07, 1.48)}, anchor=north east, /tikz/every even column/.append style={column sep=0.2em}},
legend columns=3
]
\addplot [ultra thick, color1] table {power_plots_new/optim_fc_0.07.dat};
\addplot [ultra thick, color2] table {power_plots_new/optim_fc_0.075.dat};
\addplot [ultra thick, color3] table {power_plots_new/optim_fc_0.08.dat};
\addplot [ultra thick, color4] table {power_plots_new/optim_fc_0.085.dat};
\addplot [ultra thick, color5] table {power_plots_new/optim_fc_0.09.dat};
\addplot [ultra thick, color6] table {power_plots_new/optim_fc_0.095.dat};
\addplot [ultra thick, color7] table {power_plots_new/optim_fc_0.1.dat};
\legend{$E_{\mathrm{max}}=0.07$[J], $E_{\mathrm{max}}=0.075$[J], $E_{\mathrm{max}}=0.08$[J], $E_{\mathrm{max}}=0.085$[J], $E_{\mathrm{max}}=0.09$[J], $E_{\mathrm{max}}=0.095$[J], $E_{\mathrm{max}}=0.1$[J]}

\end{axis}
\end{tikzpicture}}
    \subfloat[Latency $F_{T}^{-1}(\varrho)$. \label{fig:optim_l}]{
\begin{tikzpicture}

\definecolor{darkgray176}{RGB}{176,176,176}

\definecolor{crimson2143940}{RGB}{240, 249, 33}
\definecolor{darkorange25512714}{RGB}{253,180,47}
\definecolor{forestgreen4416044}{RGB}{237,121,83}
\definecolor{mediumpurple148103189}{RGB}{204,71,120}
\definecolor{orchid227119194}{RGB}{156, 23, 158}
\definecolor{sienna1408675}{RGB}{92,1,166}
\definecolor{steelblue31119180}{RGB}{13,8,135}

\begin{axis}[
width=24em,
height=13em,
grid=both,
grid style={line width=.1pt, draw=gray!10},
major grid style={line width=.2pt,draw=gray!40},
minor tick num=1,
xlabel={$Q$},
ylabel={$F_{T}^{-1}(\varrho)$ [s]},
x grid style={darkgray176},
xmin=1.0, xmax=1.5,
xtick style={color=black},
y grid style={darkgray176},
ymin=0.25, ymax=0.45,
ytick style={color=black},
legend style={draw=white!15!black, font=\scriptsize, at={(1.07, 1.48)}, anchor=north east, /tikz/every even column/.append style={column sep=0.2em}},
legend columns=3
]
\addplot [ultra thick, color1] table {power_plots_new/optim_quantile_0.08.dat};
\addplot [ultra thick, color2] table {power_plots_new/optim_quantile_0.085.dat};
\addplot [ultra thick, color3] table {power_plots_new/optim_quantile_0.09.dat};
\addplot [ultra thick, color4] table {power_plots_new/optim_quantile_0.095.dat};
\addplot [ultra thick, color5] table {power_plots_new/optim_quantile_0.1.dat};
\addplot [ultra thick, color6] table {power_plots_new/optim_quantile_0.11.dat};
\addplot [ultra thick, color7] table {power_plots_new/optim_quantile_0.12.dat};
\legend{$E_{\mathrm{max}}=0.07$[J], $E_{\mathrm{max}}=0.075$[J], $E_{\mathrm{max}}=0.08$[J], $E_{\mathrm{max}}=0.085$[J], $E_{\mathrm{max}}=0.09$[J], $E_{\mathrm{max}}=0.095$[J], $E_{\mathrm{max}}=0.1$[J]}

\end{axis}
\end{tikzpicture}}
    \caption{Optimal frequency and latency of the unconstrained problem \eqref{eq:unconstrained_prob} for different values of the maximum energy $E_{\mathrm{max}}$ and $\varrho=0.99$.}
    \vspace{-0.4cm}
\end{figure*}

\label{sec:results}
\AZ{In this section we present some results obtained in the two target scenarios to gain insight} 
into the interplay between communication and computation in terms of energy consumption, latency and reliability. 
We report the system parameters in Table~\ref{tab:parameters}, while the channel model and lossless compression ratio interval are taken from \cite{suman2023statistical}. For energy consumption parameters and \gls{cpu} specifications, we refer to \cite{Israel@Power_Freq}. Reproducible code for the results is provided.\footnote{https://github.com/pietro-talli/energy-qaoi-cc}
For a given set of application requirements and communication parameters, the optimization determines a joint operating point for $Q$, $f_c$, and, where applicable, $\alpha$. The solution need not be recomputed for every packet or instantaneous channel realization; it may instead be updated when parameters such as the data-block size, energy or timing budget, reliability target, or channel statistics change. In practice, these operating points may be obtained by rerunning the optimization or retrieved from a precomputed lookup table. Although different variables may be reconfigured at different time scales depending on the hardware and protocol implementation, they are jointly optimized within each problem instance.

In the following, we will mostly report results for the slow fading scenario. However, we include the main results for the fast fading scenario, and all considerations are valid for both. In that scenario, we considered a cohrence time $T_{\mathrm{coh}}\approx\frac{0.423}{f_D}$ where $f_D$ is the maximum Doppler shift \cite{rappaport2010wireless}.

\subsection{Upper Bound Approximation Error Analysis}
\label{sec:up_bound_error}
As a first validation step, we verify the accuracy of the communication-latency approximation in~\eqref{eq:channel_gauss_approx}. Following~\cite{li2025unified}, we use \gls{spa} as a numerical reference for the latency \gls{cdf}, and we benchmark it against Monte Carlo simulations of the underlying negative binomial transmission time model. We then report the approximation error as a function of the reliability target $\varrho$ and the packet error probability $\varepsilon$ in Figs.~\ref{fig:err_rel} and~\ref{fig:err_eps}, respectively. The results show close agreement in the regimes of interest; the observed error is limited (approximately $2.6\%$) even for very stringent reliability (e.g., $\varrho=0.99999$).

We further validate the upper bound in~\eqref{eq:quantile_bound} across the key parameters governing both the communication and compression latency components. Specifically, Fig.~\ref{fig:quantile_bound} reports the gap between the end-to-end latency quantile estimated via Monte Carlo simulation and the corresponding upper bound given by~\eqref{eq:quantile_bound}. Across the considered regimes, the bound is typically tight, with a gap between $0\%$ and $1.25\%$; only for the most stringent reliability values does the bound become more conservative, yielding an overestimation of up to approximately $2.5\%$. These results substantiate the accuracy of the proposed relaxation and justify its use in the subsequent optimization.
We provide an additional validation to illustrate how the true end-to-end latency quantile behaves relative to the bound. In Fig.~\ref{fig:quantile_bound_99} and~\ref{fig:quantile_bound_99999}, we show that the Monte Carlo estimate of the end-to-end latency quantile lies within a narrow band delimited by the upper bound in~\eqref{eq:quantile_bound} and $0.99$ times that bound. This shows that the bound is tight and that the quantile curve closely follows it across the design space, preserving the convex dependence used in optimization. Since the gap is small, the plotted range of $Q$ is restricted for clarity. With variability on the order of $1$--$3\%$, the approximation error remains acceptable, supporting the proposed modeling and optimization of the energy-latency trade-off.

\subsection{Energy-Constrained Scenario}
\AZ{In this scenario, we fix the maximum available energy $ E_{\mathrm{max}} $ and \AM{compute} the Pareto front, i.e., \AM{the combination of} other parameters ($Q$ and $f_c$) \AM{that minimize the latency $\varrho$-quantile while satisfying the energy constraint}. The analysis is repeated for progressively larger values of the energy constraint $ E_{\mathrm{max}} $ 
to explore how the latency quantile evolves.} 

\par  \AM{To this end, we first illustrate the relationship between the optimal \gls{cpu} frequency $f_{c}^{*}$ and $Q$, as established in Lemma~\ref{lem:eq_const}, which reduces problem~\eqref{eq:Power_Const_Opt} to a single-variable optimization in~$Q$. Fig.~\ref{fig:optim_f} illustrates the Pareto fronts for different values of \( E_{\mathrm{max}} \), showing the optimal \gls{cpu} frequency \( f_c^* \) as a function of the compression ratio \( Q \) for each energy constraint.
} 
\AZ{It is interesting to observe that the $f_{c}^{*}$ versus $Q$ curve is concave for smaller values of the energy budget, and convex for larger values. This change is due to the different rate at which the numerator and denominator of $f_c^*$ in \eqref{eq:f_c_star} change with $Q$. }\AM{For small energy budgets, the numerator of \eqref{eq:f_c_star} rises quickly at first, while the denominator grows slowly; this makes the ratio, and thus the optimal frequency, increase initially. As the numerator’s growth tapers off and the denominator’s exponential growth accelerates, the ratio decreases. For large energy budgets, the numerator is almost constant, so the exponentially increasing denominator drives a steady, convex decline in the ratio.}  \AM{Note that for certain values of \( Q \), a feasible solution for \( f_c^* \) may not exist. This infeasibility stems from hardware constraints on compression: the energy required for compression depends on the \gls{cpu}’s operating frequency and power consumption, both of which are bounded by physical limits. In practice, processors cannot operate below a minimum frequency \( f_{c,\min} \) nor above a maximum frequency \( f_{c,\max} \) due to design and thermal constraints. Consequently, we restrict the feasible range  to \( 800\,\mathrm{MHz} \leq f_c^* \leq 2.5\,\mathrm{GHz} \), which reflects realistic hardware capabilities \cite{capra2019edge}.}

Next, \AZ{in Fig.~\ref{fig:optim_l} we report the latency $\varrho$-quantile,  denoted by the inverse \gls{cdf} $ F_T^{-1} $. \PT{This corresponds to the minimal latency that still satisfies the reliability requirement $\varrho$ and is}
\AM{obtained using the same $(f_c^*, Q)$ parameter pairs that generate the Pareto fronts in Fig.~\ref{fig:optim_f}}. We observe that the latency profiles turn out to be  convex, thus \AM{ indicating that for each  $E_{\mathrm{max}}$, there exists a compression ratio $Q$ that minimizes the latency quantile}.} 
\AZ{Comparing the two figures, \AM{i.e. Fig.~\ref{fig:optim_f} and Fig.~\ref{fig:optim_l}},} a general trend emerges: as the available energy $ E_{\mathrm{max}} $ increases, the optimal operating frequency $ f_c^* $ also increases, leading to a reduction in system latency. This confirms the \AZ{intuition} that higher energy budgets enable faster processing and more efficient compression, thereby improving overall performance.

\begin{figure}
    \centering
\begin{tikzpicture}

\definecolor{darkgray176}{RGB}{176,176,176}

\definecolor{crimson2143940}{RGB}{240, 249, 33}
\definecolor{darkorange25512714}{RGB}{253,180,47}
\definecolor{forestgreen4416044}{RGB}{237,121,83}
\definecolor{mediumpurple148103189}{RGB}{204,71,120}
\definecolor{orchid227119194}{RGB}{156, 23, 158}
\definecolor{sienna1408675}{RGB}{92,1,166}
\definecolor{steelblue31119180}{RGB}{13,8,135}

\begin{axis}[
width=25em,
height=15.5em,
grid=both,
grid style={line width=.1pt, draw=gray!10},
major grid style={line width=.2pt,draw=gray!40},
minor tick num=1,
xlabel={Latency [s]},
ylabel={Energy [J]},
x grid style={darkgray176},
xmin=0.2, xmax=0.6,
xtick style={color=black},
y grid style={darkgray176},
ymin=0.06, ymax=0.14,
ytick style={color=black},
ytick={0.06,0.08, 0.1, 0.12, 0.14},
yticklabels={0.06, 0.08, 0.1, 0.12, 0.14},
legend style={draw=white!15!black, font=\scriptsize, at={(0.99, 0.98)}, anchor=north east, /tikz/every even column/.append style={column sep=0.2em}},
legend columns=1
]
\addplot [ultra thick, color2]
table[
    x=Quantile,
    y=Energy,
    col sep=comma
] {figures/slow_fading/first_scenario/pareto_rho_0.9.txt};

\addplot [ultra thick, color3]
table[
    x=Quantile,
    y=Energy,
    col sep=comma
] {figures/slow_fading/first_scenario/pareto_rho_0.99.txt};

\addplot [ultra thick, color4]
table[
    x=Quantile,
    y=Energy,
    col sep=comma
] {figures/slow_fading/first_scenario/pareto_rho_0.999.txt};

\addplot [ultra thick, color5]
table[
    x=Quantile,
    y=Energy,
    col sep=comma
] {figures/slow_fading/first_scenario/pareto_rho_0.9999.txt};

\addplot [ultra thick, color6]
table[
    x=Quantile,
    y=Energy,
    col sep=comma
] {figures/slow_fading/first_scenario/pareto_rho_0.99999.txt};

\addplot [semithick, color7, only marks, mark=*, mark size=2, mark options={solid}]
table {%
0.3105219988336671 0.07698759782619169
};

\addlegendimage{only marks, thick,  mark=triangle}

\addplot [black, dashed, name path=low] table {%
0.31052199883366711 0.07698759782619169
0.7   0.07698759782619169
};
\addplot [black, name path=high] table {%
0.3105219988336671 0.14
0.7   0.14
};
\addplot [black, dashed, name path=left] table {%
0.3105219988336671 0.07698759782619169
0.3105219988336671 0.14
};
\addplot[
        fill=black!5,
        draw=none,
        pattern=north west lines,
        pattern color=black
    ] fill between[
        of=high and low
    ];

\addplot[thick, color6, mark=triangle] coordinates {(3.055144600151617484e-01,8.131313131313130937e-02)};
\addplot[thick, color5, mark=triangle] coordinates {(2.982081039769094444e-01,8.535353535353534804e-02)};
\addplot[thick, color4, mark=triangle] coordinates {(2.882202941285338382e-01,8.939393939393938671e-02)};
\addplot[thick, color3, mark=triangle] coordinates {(2.735276269466385690e-01,1.035353535353535359e-01)};
\addplot[thick, color2, mark=triangle] coordinates {(2.489831750235970143e-01,1.318181818181818343e-01)};

\legend{$\varrho=0.9$, $\varrho=0.99$, $\varrho=0.999$, $\varrho=0.9999$,  $\varrho=0.99999$, No compression, Energy-unaware}
\end{axis}
\end{tikzpicture}
    \vspace{-0.6cm}
    \caption{Pareto front for the energy-constrained scenario.}
    \label{fig:results_scen1}
    \vspace{-0.4cm}
\end{figure}

\begin{figure}
    \centering
\begin{tikzpicture}

\definecolor{darkgray176}{RGB}{176,176,176}

\definecolor{crimson2143940}{RGB}{240, 249, 33}
\definecolor{darkorange25512714}{RGB}{253,180,47}
\definecolor{forestgreen4416044}{RGB}{237,121,83}
\definecolor{mediumpurple148103189}{RGB}{204,71,120}
\definecolor{orchid227119194}{RGB}{156, 23, 158}
\definecolor{sienna1408675}{RGB}{92,1,166}
\definecolor{steelblue31119180}{RGB}{13,8,135}

\begin{axis}[
width=25em,
height=15.5em,
grid=both,
grid style={line width=.1pt, draw=gray!10},
major grid style={line width=.2pt,draw=gray!40},
minor tick num=1,
xlabel={Latency [s]},
ylabel={Energy [J]},
x grid style={darkgray176},
xmin=0.2, xmax=0.6,
xtick style={color=black},
y grid style={darkgray176},
ymin=0.06, ymax=0.14,
ytick style={color=black},
ytick={0.06,0.08, 0.1, 0.12, 0.14},
yticklabels={0.06, 0.08, 0.1, 0.12, 0.14},
legend style={draw=white!15!black, font=\scriptsize, at={(0.99, 0.98)}, anchor=north east, /tikz/every even column/.append style={column sep=0.2em}},
legend columns=1
]
\addplot [ultra thick, color2]
table[
    x=Quantile,
    y=Energy,
    col sep=comma
] {figures/fast_fading/first_scenario/pareto_rho_0.9.txt};

\addplot [ultra thick, color3]
table[
    x=Quantile,
    y=Energy,
    col sep=comma
] {figures/fast_fading/first_scenario/pareto_rho_0.99.txt};

\addplot [ultra thick, color4]
table[
    x=Quantile,
    y=Energy,
    col sep=comma
] {figures/fast_fading/first_scenario/pareto_rho_0.999.txt};

\addplot [ultra thick, color5]
table[
    x=Quantile,
    y=Energy,
    col sep=comma
] {figures/fast_fading/first_scenario/pareto_rho_0.9999.txt};

\addplot [ultra thick, color6]
table[
    x=Quantile,
    y=Energy,
    col sep=comma
] {figures/fast_fading/first_scenario/pareto_rho_0.99999.txt};

\addplot [semithick, color7, only marks, mark=*, mark size=2, mark options={solid}]
table {%
0.3175087348577471 0.0772309013290442
};

\addlegendimage{only marks, thick,  mark=triangle}

\addplot [black, dashed, name path=low] table {%
0.3175087348577471 0.0772309013290442
0.7   0.0772309013290442
};
\addplot [black, name path=high] table {%
0.3175087348577471 0.14
0.7   0.14
};
\addplot [black, dashed, name path=left] table {%
0.3175087348577471 0.0772309013290442
0.3175087348577471 0.14
};
\addplot[
        fill=black!5,
        draw=none,
        pattern=north west lines,
        pattern color=black
    ] fill between[
        of=high and low
    ];

\addplot[thick, color6, mark=triangle] coordinates {(3.062966311355379001e-01,8.131313131313130937e-02)};
\addplot[thick, color5, mark=triangle] coordinates {(2.992096817221839289e-01, 8.535353535353534804e-02)};
\addplot[thick, color4, mark=triangle] coordinates {(0.2890289395607174017, 0.09141414141414141992)};
\addplot[thick, color3, mark=triangle] coordinates {(2.743615975117160155e-01, 1.025252525252525193e-01)};
\addplot[thick, color2, mark=triangle] coordinates {(2.499186029427775402e-01, 1.338383838383838398e-01)};

\legend{$\varrho=0.9$, $\varrho=0.99$, $\varrho=0.999$, $\varrho=0.9999$,  $\varrho=0.99999$, No compression, Energy-unaware}
\end{axis}
\end{tikzpicture}
    \vspace{-0.6cm}
    \caption{Pareto front for the energy-constrained scenario (fast fading channel).}
    \label{fig:fs_ff}
    \vspace{-0.4cm}
\end{figure}

\AM{Fig.~\ref{fig:results_scen1} illustrates the Pareto front} \AZ{in terms of Energy and Latency $\varrho$-quantile. Higher $\varrho$ values correspond to progressively stricter maximum-latency guarantees, revealing how tightening reliability requirements reshapes the achievable trade-off.}
\AM{A Pareto front for a given latency quantile or reliability level $\varrho$ is obtained by computing, for each energy constraint, the minimum achievable latency. Equivalently, each point on the front corresponds to a unique combination of \gls{cpu} frequency \( f_c \) and compression ratio \( Q \). For comparison, the system’s baseline configuration, operating without compression, is marked by a colored dot whenever it lies on the front.} \AM{The results show that, at lower reliability levels, the joint optimization of \( f_c \) and \( Q \) consistently yields Pareto-superior configurations compared to the uncompressed baseline. In this regime, compression reduces the data volume to be transmitted, lowering both energy consumption and total latency without incurring a significant penalty from variability in transmission times. The steepness of the Pareto curves in this region highlights that substantial energy savings can be achieved with only marginal increases in latency, underscoring the benefits of jointly optimizing compression and communication parameters when energy budgets are limited. At higher reliability levels, however, reducing the available energy budget causes latency to increase significantly after a certain point. This is because the additional processing time required for compression becomes more significant, particularly in the worst-case tail of the latency distribution. In this setting, we find that the minimum achievable latency, obtained without compression, can be matched at an equal or a lower overall energy budget than with compression, making transmission without compression the more favorable choice.}

We also included the uncompressed case where no compression is used. This is represented as single point in the energy-latency plot in Fig.~\ref{fig:results_scen1}. We also represent the suboptimal region with respect to this point with a black pattern to enforce that those point should be excluded from any Pareto front as they are sub-optimal. We can see that some points of the $\varrho=0.9999$ and $\varrho=0.99999$ are inside this sub-optimal region. This means that in these cases any possible combination of compression and communication gives a worse trade-off. With such reliability values, improvements with respect to the no compression case are seen only for higher latencies, where it is possible to achieve lower energy consumption. The plot also shows the \textit{energy-unaware} result, which shows the minimum latency obtained when we remove the energy constraint. We mark the corresponding point with a triangle for each reliability $\varrho$. With the given parameters, increasing the energy would not allow any further gain in terms of latency guarantees.

The results for the fast fading scenario are reported in Fig.~\ref{fig:fs_ff}. We can note that the performance of the system is only slightly lower, as the achievable rate is extremely similar. In particular, the achievable latency is slightly higher for the same energy and reliability requirements, but the difference is almost negligible.

We also considered the fixed compression ratio scenario. For fixed values of the compression ratios, the optimal $f_c^*$ is computed and different energy and latencies values are obtained by varying the energy constraint $E_{\mathrm{max}}$. In Fig.~\ref{fig:comp_fixed} it is shown a  comparison between fixed compression ratios and the optimal Pareto front for $\varrho=0.99$. It can be observed that while low $Q$ values are highly suboptimal, for larger values of $Q$ the gap with the optimal achievable trade-off is smaller. We further analyze the compression energy as a function of the optimal $f_c^*$ and $Q^*$ to assess its impact on the trade-off. It is important to note that its contribution becomes appreciable only under stringent low-latency requirements.

\begin{figure}
    \centering
\begin{tikzpicture}

\definecolor{darkgray176}{RGB}{176,176,176}

\definecolor{crimson2143940}{RGB}{240, 249, 33}
\definecolor{darkorange25512714}{RGB}{253,180,47}
\definecolor{forestgreen4416044}{RGB}{237,121,83}
\definecolor{mediumpurple148103189}{RGB}{204,71,120}
\definecolor{orchid227119194}{RGB}{156, 23, 158}
\definecolor{sienna1408675}{RGB}{92,1,166}
\definecolor{steelblue31119180}{RGB}{13,8,135}

\begin{axis}[
width=25em,
height=14.5em,
grid=both,
grid style={line width=.1pt, draw=gray!10},
major grid style={line width=.2pt,draw=gray!40},
minor tick num=1,
xlabel={Latency [s]},
ylabel={Energy [J]},
x grid style={darkgray176},
xmin=0.27, xmax=0.37,
xtick style={color=black},
y grid style={darkgray176},
ymin=0.0, ymax=0.14,
ytick style={color=black},
ytick={0, 0.02, 0.04, 0.06, 0.08, 0.1, 0.12, 0.14},
yticklabels={0, 0.02, 0.04, 0.06, 0.08, 0.1, 0.12, 0.14},
legend style={draw=white!15!black, font=\scriptsize, at={(0.5, 1.28)}, anchor=north, /tikz/every even column/.append style={column sep=0.2em}},
legend columns=3
]

\addplot [ultra thick, color3]
table[
    x=Quantile,
    y=Energy,
    col sep=comma
] {figures/fast_fading/first_scenario/pareto_rho_0.99.txt};

\addplot[thick, color3, dashed] table {
0.37182912365807264 0.005694161122457277
0.34901017749712526 0.00454231107590676
0.335164113779994 0.003937758325056606
0.3247422365344312 0.004228295430666551
0.31634265901789704 0.004251787967744114
0.3096212874862495 0.004756503850545622
0.30430932829046986 0.00525290065693701
0.30011003813968584 0.0057407712734932614
0.2968004423551807 0.006486493154940487
0.2941577671784557 0.00723000266424062
0.2920078322612524 0.008240103674341637
0.29022045484573405 0.00925020468444264
0.2886971783239336 0.010526897195344526
0.28737253725109835 0.011536998205445528
0.2862047119983673 0.012811478344200336
0.28515637239857783 0.013821579354301339
0.28420705577099087 0.01509387454694272
0.28333970906883504 0.016103975557043737
0.28253737275468305 0.01737411277814914
0.2817964457618856 0.018384213788250142
0.28110002122803046 0.019652219570191892
0.28045164972395087 0.02066232058029291
0.2798358845598629 0.021928221022262794
0.27925775243393275 0.022938322032363796
0.27870629495863614 0.024202142809177533
0.27818299232941646 0.02521224381927855
0.27768482072679385 0.026474010189966327
0.2772053008235938 0.02748411120006733
0.2767515104332489 0.028494212210168346
0.27630884393706984 0.02975394902635771
0.27589643623725135 0.0310116807404541
0.2755137331137171 0.032267431413625494
0.2752509610108957 0.03376298296215206
0.2749500749969592 0.03501293108297569
0.2748540393161585 0.036497082866239136
0.2746362886322831 0.03774143162811734
0.27444795005699735 0.038983957454506335
0.27457998363385683 0.04045352375085352
0.2744772170084052 0.04169070664886679
0.2744043897339987 0.04292614945974514
0.274361597511716 0.04415987210810524
};

\addplot [semithick, color7, only marks, mark=*, mark size=2, mark options={solid}]
table {%
0.3175087348577471 0.0772309013290442
};

\addplot[mark = square, color4] table {
0.3084939327163582 0.07121212121212121
0.3016695515330936 0.07222222222222223
0.29865777755991607 0.07323232323232323
0.29686502334219955 0.07424242424242425
0.29564254611981866 0.07525252525252525
0.2947405937528884 0.07626262626262627
0.29403983298696623 0.07727272727272727
0.2934751048926007 0.07828282828282829
0.2930074200149532 0.0792929292929293
};

\addplot[mark = diamond, color5] table {
0.33601075458998725 0.06414141414141414
0.327319984012434 0.06515151515151515
0.32066067433953005 0.06616161616161617
0.31534759617485997 0.06717171717171717
0.31098080228590297 0.06818181818181818
0.30730909246938265 0.0691919191919192
0.30416569394374654 0.07020202020202021
0.3014349642502957 0.07121212121212121
0.2990338549002554 0.07222222222222223
0.2969009777407375 0.07323232323232323
0.2949898438876667 0.07424242424242425
0.29326451514467855 0.07525252525252525
0.29169671240547584 0.07626262626262627
0.29026383717680715 0.07727272727272727
0.2889475839102759 0.07828282828282829
0.28773294535528804 0.0792929292929293
0.2866074858114804 0.0803030303030303
0.2855608009786782 0.08131313131313131
0.28458411030412145 0.08232323232323233
0.2836699450536212 0.08333333333333334
0.28281190662810607 0.08434343434343435
0.28200447716454236 0.08535353535353535
0.28124286955893435 0.08636363636363636
0.28052290756717135 0.08737373737373738
0.27984092910531677 0.08838383838383838
0.27919370762433177 0.08939393939393939
0.27857838769769394 0.0904040404040404
0.27799243188204564 0.09141414141414142
0.27743357659104195 0.09242424242424244
0.2768997952296587 0.09343434343434344
0.2763892672180894 0.09444444444444444
0.27590035182460126 0.09545454545454546
0.27543156594922386 0.09646464646464648
0.27498156517211464 0.09747474747474748
0.27454912751437116 0.09848484848484848
};

\addplot[mark = triangle, color6] table {
0.4513545206105338 0.062121212121212126
0.4406384020659849 0.06313131313131314
0.43123248306942275 0.06414141414141414
0.4228898425679431 0.06515151515151515
0.41542415377078334 0.06616161616161617
0.4086918153141441 0.06717171717171717
0.4025801276495642 0.06818181818181818
0.39699923738610693 0.0691919191919192
0.391876508047021 0.07020202020202021
0.3871524982628529 0.07121212121212121
0.3827780316621371 0.07222222222222223
0.37871202468606613 0.07323232323232323
0.37491985102019204 0.07424242424242425
0.3713720926973456 0.07525252525252525
0.3680435742816329 0.07626262626262627
0.3649126072994042 0.07727272727272727
0.3619603928835705 0.07828282828282829
0.3591705449123296 0.0792929292929293
0.3565287059326314 0.0803030303030303
0.3540222352604596 0.08131313131313131
0.3516399537570283 0.08232323232323233
0.34937193349846674 0.08333333333333334
0.34720932329538956 0.08434343434343435
0.345144203057705 0.08535353535353535
0.3431694615331716 0.08636363636363636
0.34127869311181935 0.08737373737373738
0.3394661102792319 0.08838383838383838
0.33772646898936004 0.08939393939393939
0.3360550047624572 0.0904040404040404
0.33444737773285615 0.09141414141414142
0.33289962520195543 0.09242424242424244
0.33140812051432955 0.09343434343434344
0.3299695372846249 0.09444444444444444
0.3285808181714439 0.09545454545454546
0.3272391475305976 0.09646464646464648
0.32594192739071737 0.09747474747474748
0.3246867562845055 0.09848484848484848
0.3234714105429526 0.0994949494949495
0.3222938277208677 0.10050505050505051
0.32115209187254756 0.10151515151515152
0.3200444204383638 0.10252525252525252
0.3189691525380405 0.10353535353535354
0.31792473849569763 0.10454545454545455
0.31690973044636433 0.10555555555555556
0.315922773894436 0.10656565656565657
0.31496260011212573 0.10757575757575757
0.31402801928088064 0.10858585858585859
0.3131179142914426 0.10959595959595961
0.31223123512908213 0.11060606060606061
0.311366993779835 0.11161616161616161
0.31052425960155283 0.11262626262626263
0.30970215511045907 0.11363636363636363
0.30889985213984006 0.11464646464646465
0.30811656833264034 0.11565656565656567
0.3073515639341956 0.11666666666666667
0.30660413885521276 0.11767676767676769
0.3058736299784933 0.11868686868686869
0.30515940868584757 0.1196969696969697
0.3044608785842369 0.12070707070707071
0.3037774734124508 0.12171717171717172
0.30310865511161766 0.12272727272727273
0.30245391204461036 0.12373737373737374
0.3018127573509543 0.12474747474747475
0.3011847274252172 0.12575757575757576
0.30056938050807536 0.12676767676767675
0.29996629538032527 0.12777777777777777
0.2993750701510689 0.12878787878787878
0.29879532113214957 0.1297979797979798
0.2982266817916773 0.13080808080808082
0.2976688017801578 0.13181818181818183
0.29712134602334705 0.13282828282828285
0.2965839938764936 0.13383838383838384
0.2960564383351202 0.13484848484848486
0.2955383852979306 0.13585858585858585
0.2950295528778213 0.13686868686868686
0.2945296707573315 0.13787878787878788
0.29403847958518353 0.1388888888888889
0.2935557304108524 0.1398989898989899
0.2930811841543658 0.14090909090909093
0.2926146111087678 0.14191919191919192
0.2921557904728954 0.14292929292929293
0.29170450991230923 0.14393939393939392
0.29126056514639415 0.14494949494949494
0.29082375955980744 0.14595959595959596
0.29039390383659497 0.14696969696969697
0.2899708156154288 0.147979797979798
};

\addplot [black, dashed, name path=low] table {%
0.3175087348577471 0.0772309013290442
0.7   0.0772309013290442
};
\addplot [black, name path=high] table {%
0.3175087348577471 0.14
0.7   0.14
};
\addplot [black, dashed, name path=left] table {%
0.3175087348577471 0.0772309013290442
0.3175087348577471 0.14
};
\addplot[
        fill=black!5,
        draw=none,
        pattern=north west lines,
        pattern color=black
    ] fill between[
        of=high and low
    ];

\legend{Optimal, {$E_c(f_c^*,Q^*)$}, No compression, $Q=1.1$,$Q=1.3$,$Q=1.5$};

\end{axis}
\end{tikzpicture}\vspace{-0.5cm}
    \caption{Comparison between fixed compression and optimal trade-off for $\varrho=0.99$.}
    \label{fig:comp_fixed}
    \vspace{-0.4cm}
\end{figure}

\subsection{Time-Constrained Scenario}
\AM{In the time-constrained scenario, we fix the value of \( T \) (or equivalently the \gls{qaoi}) and determine the corresponding Pareto front, representing the set of \((f_c, Q)\) combinations that minimize the total energy consumed in both compression and communication, while satisfying the overall reliability requirement. \pp{} In this setting, the three optimization variables; the compression/transmission allocation factor \( \alpha \), the \gls{cpu} frequency \( f_c \), and the compression ratio \( Q \); are tightly coupled and not explicitly separable. To provide deeper insight into their individual roles and mutual dependencies, we first present a series of plots that illustrate how variations in these parameters influence both the objective function and the overall reliability constraint in problem~\eqref{eq:Time_Const_opt}.} Specifically, we examine how the overall reliability or success probability, $P_{\mathrm{succ}}= (1 - \varepsilon_c)(1 - \varepsilon_{\mathrm{tx}})$, and the total energy consumption, \( E_c + E_{\mathrm{tx}} \), vary with respect to each optimization variable, while keeping the other two fixed. In each plot, the two curves correspond to different values of \( T \), thereby highlighting the sensitivity of system performance to \gls{qaoi} constraints.

\begin{figure}[!t]
    \centering
    \subfloat{
\begin{tikzpicture}

\begin{axis}[
    width=0cm,
    height=0cm,
    axis line style={draw=none},
    tick style={draw=none},
    at={(0,0)},
    scale only axis,
    xmin=0,
    xmax=0,
    xtick={},
    ymin=0,
    ymax=0,
    ytick={},
    axis background/.style={fill=white},
    legend style={legend cell align=left, fill opacity=1, align=center, draw=black, font=\scriptsize, at={(0, 0)}, anchor=center, /tikz/every even column/.append style={column sep=0.3em}},
    legend columns=2,
]
\addplot [ultra thick, color7, const plot mark left]
table {%
0 0
};
\addlegendentry{Level A}
\addplot [ultra thick, color2, const plot mark left]
table {%
0 0
};
\addlegendentry{Level B}

\legend{$T=0.4$s, $T=0.5$s};
\end{axis}
\end{tikzpicture}} \\ 
    \setcounter{subfigure}{0}
\begin{tikzpicture}

\definecolor{darkgray176}{RGB}{176,176,176}

\definecolor{crimson2143940}{RGB}{240, 249, 33}
\definecolor{darkorange25512714}{RGB}{253,180,47}
\definecolor{forestgreen4416044}{RGB}{237,121,83}
\definecolor{mediumpurple148103189}{RGB}{204,71,120}
\definecolor{orchid227119194}{RGB}{156, 23, 158}
\definecolor{sienna1408675}{RGB}{92,1,166}
\definecolor{steelblue31119180}{RGB}{13,8,135}

\begin{axis}[
width=11em,
height=11em,
grid=both,
grid style={line width=.1pt, draw=gray!10},
major grid style={line width=.2pt,draw=gray!40},
minor tick num=1,
xlabel={$\alpha$},
ylabel={$P_{\rm succ}$},
x grid style={darkgray176},
xmin=0.0, xmax=1.0,
xtick style={color=black},
y grid style={darkgray176},
ymin=0.0, ymax=1,
ytick style={color=black},
legend style={draw=white!15!black, font=\scriptsize, at={(0.99, 0.98)}, anchor=north east, /tikz/every even column/.append style={column sep=0.2em}},
legend columns=1
]
\addplot [ultra thick, color2] table {qoai_plots/2_prob_alpha_Q1.20_fc1600000000.00.dat};
\addplot [ultra thick, color7] table {qoai_plots/prob_alpha_Q1.20_fc1600000000.00.dat};

\end{axis}
\end{tikzpicture}
    \hspace{-1em}
\pgfplotsset{scaled y ticks=false}
\begin{tikzpicture}

\definecolor{darkgray176}{RGB}{176,176,176}

\definecolor{crimson2143940}{RGB}{240, 249, 33}
\definecolor{darkorange25512714}{RGB}{253,180,47}
\definecolor{forestgreen4416044}{RGB}{237,121,83}
\definecolor{mediumpurple148103189}{RGB}{204,71,120}
\definecolor{orchid227119194}{RGB}{156, 23, 158}
\definecolor{sienna1408675}{RGB}{92,1,166}
\definecolor{steelblue31119180}{RGB}{13,8,135}

\begin{axis}[
width=11em,
height=11em,
grid=both,
grid style={line width=.1pt, draw=gray!10},
major grid style={line width=.2pt,draw=gray!40},
minor tick num=1,
xlabel={$\alpha$},
ylabel={$E_c + E_{\mathrm{tx}}$[J]},
x grid style={darkgray176},
xmin=0.0, xmax=1.0,
xtick style={color=black},
y grid style={darkgray176},
ymin=0.0, ymax=0.15,
ytick style={color=black},
ytick={0, 0.05,0.1,0.15},
yticklabels={0.0, 0.05, 0.1,0.15},
legend style={draw=white!15!black, font=\scriptsize, at={(0.99, 0.98)}, anchor=north east, /tikz/every even column/.append style={column sep=0.2em}},
legend columns=1
]

\addplot [ultra thick, color2] table {qaoi_plots_new/2_energy_alpha_Q1.20_fc1600000000.00.dat};

\addplot [ultra thick, color7] table {qaoi_plots_new/energy_alpha_Q1.20_fc1600000000.00.dat};

\end{axis}
\end{tikzpicture}
    \caption{Overall success probability and total energy consumption for different values of $\alpha$, $Q=1.2$ and $f_c=1.6$ [GHz].}
    \label{fig:alpha_prob}
\end{figure}
\par \AM{We begin with Fig.~\ref{fig:alpha_prob}, which examines the influence of $ \alpha $, the fraction of the available time slot allocated to compression}. The \AM{left} plot shows the value of $P_{\rm succ} $ for fixed $ Q $ and $ f_c $ with varying $\alpha$. For very small values of $ \alpha $, the communication phase dominates the time allocation, leaving insufficient time for compression. \AM{In this regime, the compression success probability $ (1 - \varepsilon_{\mathrm{c}}) $ is significantly low, driving the overall reliability toward zero.} As $ \alpha $ increases, both compression and transmission phases receive adequate time, and the success probabilities improve. The product of the two probabilities approaches one, indicating high overall reliability. \AM{However, beyond a certain threshold, allocating more time to compression reduces the time available for transmission , sharply decreasing $ (1 - \varepsilon_{\mathrm{tx}}) $, and, consequently, the overall reliability drops again toward zero. Furthermore, increasing $T$ from $0.4$ to $0.5$ broadens the feasible range of $\alpha$ (for fixed $f_{c}$ and $Q$) over which the success probability remains high.}

In terms of energy, only the transmission energy $ E_{\mathrm{tx}} $ varies with $ \alpha $, since compression energy is independent of the time allocation for fixed values of $f_{c}$ and $Q$ \AZ{(provided that it is completed within the allocated slot)}. \AM{As shown in the right plot of Fig.~\ref{fig:alpha_prob}, the total energy consumption exhibits an inverse dependence on $\alpha$. For small $\alpha$, insufficient compression time reduces compression success probability, leaving more bits to be transmitted, which drives transmission energy $E_{\mathrm{tx}}$ high. As $\alpha$ increases, more bits are successfully compressed, easing the transmission burden and lowering $E_{\mathrm{tx}}$. Beyond a certain threshold, however, the available transmission time $(1-\alpha)T$ becomes too short to deliver the compressed payload. In this regime, the transmission energy contribution effectively vanishes, and the total energy drops toward the compression cost alone, which in this case is negligibly low for the given $Q$ and $f_c$. Finally, it is interesting to note that in this scenario increasing $T$ leads to higher overall energy consumption. This is because the communication slot is fully utilized to maximize reliability; without joint optimization, a larger $T$ simply results in more transmissions and hence greater energy usage. Relaxing $T$ alone is not enough in the time-constrained scenario, and energy minimization requires joint optimization of the compression and communication parameters.}

\par Next, Fig.~\ref{fig:fc_prob} illustrates the impact of \gls{cpu} frequency $ f_c $ on system performance, showing both the success probability and total energy consumption. As expected, increasing $ f_c $ enhances the probability of completing compression within the allocated time, but also increases energy consumption due to the superlinear dependence of computational energy on clock frequency. \AM{Moreover, for fixed $\alpha$ and $Q$, a larger $T$ directly translates into a longer available compression slot $\alpha T$, which improves the likelihood of completing compression successfully and thus raises the overall success probability. By contrast, the transmission energy $E_{\mathrm{tx}}$ remains constant with respect to $f_c$ and dominates the compression energy $E_c$, especially for larger $T$. } Nevertheless, the monotonic behavior of $P_{\mathrm{succ}}$ and $E_c+E_{\mathrm{tx}}$ shown in Fig.~\ref{fig:fc_prob} confirms the existence of an optimal operating point $ f_c^* $ that satisfies the reliability constraint with equality, as discussed in Section\ref{sec:scen2}.

\begin{figure}[t]
    \centering
    \subfloat{
\begin{tikzpicture}

\begin{axis}[
    width=0cm,
    height=0cm,
    axis line style={draw=none},
    tick style={draw=none},
    at={(0,0)},
    scale only axis,
    xmin=0,
    xmax=0,
    xtick={},
    ymin=0,
    ymax=0,
    ytick={},
    axis background/.style={fill=white},
    legend style={legend cell align=left, fill opacity=1, align=center, draw=black, font=\scriptsize, at={(0, 0)}, anchor=center, /tikz/every even column/.append style={column sep=0.3em}},
    legend columns=2,
]
\addplot [ultra thick, color7, const plot mark left]
table {%
0 0
};
\addlegendentry{Level A}
\addplot [ultra thick, color2, const plot mark left]
table {%
0 0
};
\addlegendentry{Level B}

\legend{$T=0.4$s, $T=0.5$s};
\end{axis}
\end{tikzpicture}} \\ 
    \setcounter{subfigure}{0}
\begin{tikzpicture}

\definecolor{darkgray176}{RGB}{176,176,176}

\definecolor{crimson2143940}{RGB}{240, 249, 33}
\definecolor{darkorange25512714}{RGB}{253,180,47}
\definecolor{forestgreen4416044}{RGB}{237,121,83}
\definecolor{mediumpurple148103189}{RGB}{204,71,120}
\definecolor{orchid227119194}{RGB}{156, 23, 158}
\definecolor{sienna1408675}{RGB}{92,1,166}
\definecolor{steelblue31119180}{RGB}{13,8,135}

\begin{axis}[
width=11em,
height=11em,
grid=both,
grid style={line width=.1pt, draw=gray!10},
major grid style={line width=.2pt,draw=gray!40},
minor tick num=1,
xlabel={$f_c$ [GHz]},
ylabel={$P_{\mathrm{succ}}$},
x grid style={darkgray176},
xmin=0.8, xmax=2.5,
xtick style={color=black},
y grid style={darkgray176},
ymin=0.9996, ymax=1,
ytick style={color=black},
ytick={0.9996,0.9997, 0.9998, 0.9999, 1.0},
yticklabels={0.9996,0.9997, 0.9998, 0.9999, 1},
legend style={draw=white!15!black, font=\scriptsize, at={(0.99, 0.98)}, anchor=north east, /tikz/every even column/.append style={column sep=0.2em}},
legend columns=1
]

\addplot [ultra thick, color7] table {qoai_plots/prob_fc_Q1.20_alpha0.50.dat};
\addplot [ultra thick, color2] table {qoai_plots/2_prob_fc_Q1.20_alpha0.30.dat};

\end{axis}
\end{tikzpicture}
\pgfplotsset{scaled y ticks=false}
\begin{tikzpicture}

\definecolor{darkgray176}{RGB}{176,176,176}

\definecolor{crimson2143940}{RGB}{240, 249, 33}
\definecolor{darkorange25512714}{RGB}{253,180,47}
\definecolor{forestgreen4416044}{RGB}{237,121,83}
\definecolor{mediumpurple148103189}{RGB}{204,71,120}
\definecolor{orchid227119194}{RGB}{156, 23, 158}
\definecolor{sienna1408675}{RGB}{92,1,166}
\definecolor{steelblue31119180}{RGB}{13,8,135}

\begin{axis}[
width=11em,
height=11em,
grid=both,
grid style={line width=.1pt, draw=gray!10},
major grid style={line width=.2pt,draw=gray!40},
minor tick num=1,
xlabel={$f_c$ [GHz]},
ylabel={$E_{c}+E_{\mathrm{tx}}$ [J]},
x grid style={darkgray176},
xmin=0.8, xmax=2.5,
xtick style={color=black},
ytick={0.07,0.08,0.09},
yticklabels={0.07,0.08,0.09},
y grid style={darkgray176},
ymin=0.07, ymax=0.09,
ytick style={color=black},
legend style={draw=white!15!black, font=\scriptsize, at={(0.99, 0.98)}, anchor=north east, /tikz/every even column/.append style={column sep=0.2em}},
legend columns=1
]

\addplot [ultra thick, color2] table {qaoi_plots_new/2_energy_fc_Q1.20_alpha0.30.dat};

\addplot [ultra thick, color7] table {qaoi_plots_new/energy_fc_Q1.20_alpha0.30.dat};

\end{axis}
\end{tikzpicture}
    \caption{Overall success probability and total energy consumption for different values of $f_c$, $Q=1.2$ and $\alpha=0.3$.}
    \label{fig:fc_prob}
    \vspace{-0.4cm}
\end{figure}

\AM{Fig.~\ref{fig:q_prob} illustrates the effect of the compression ratio $Q$ on system performance for fixed $\alpha=0.35$ and $f_c=1.6$ [GHz]. When $Q$ is very low, the compression is insufficient to reduce the data load for transmission, so the transmission success probability $(1 - \varepsilon_{\mathrm{tx}})$, and consequently $P_{\rm succ}$, remains near zero. As $Q$ increases, enough redundancy is removed to allow reliable transmission within the available transmission slot, and $P_{\rm succ}$ rises sharply. When $Q$ becomes too large, however, the compression operation itself becomes more complex and time-consuming, lowering the compression success probability $(1 - \varepsilon_c)$ and reducing $P_{\rm succ}$ again. This interaction produces the characteristic rectified-S shape of the curve.}

\begin{figure}[!t]
    \centering
    \subfloat{
\begin{tikzpicture}

\begin{axis}[
    width=0cm,
    height=0cm,
    axis line style={draw=none},
    tick style={draw=none},
    at={(0,0)},
    scale only axis,
    xmin=0,
    xmax=0,
    xtick={},
    ymin=0,
    ymax=0,
    ytick={},
    axis background/.style={fill=white},
    legend style={legend cell align=left, fill opacity=1, align=center, draw=black, font=\scriptsize, at={(0, 0)}, anchor=center, /tikz/every even column/.append style={column sep=0.3em}},
    legend columns=2,
]
\addplot [ultra thick, color7, const plot mark left]
table {%
0 0
};
\addlegendentry{Level A}
\addplot [ultra thick, color2, const plot mark left]
table {%
0 0
};
\addlegendentry{Level B}

\legend{$T=0.4$s, $T=0.5$s};
\end{axis}
\end{tikzpicture}} \\ 
    \vspace{0.1cm}
    \setcounter{subfigure}{0}
\begin{tikzpicture}

\definecolor{darkgray176}{RGB}{176,176,176}

\definecolor{crimson2143940}{RGB}{240, 249, 33}
\definecolor{darkorange25512714}{RGB}{253,180,47}
\definecolor{forestgreen4416044}{RGB}{237,121,83}
\definecolor{mediumpurple148103189}{RGB}{204,71,120}
\definecolor{orchid227119194}{RGB}{156, 23, 158}
\definecolor{sienna1408675}{RGB}{92,1,166}
\definecolor{steelblue31119180}{RGB}{13,8,135}

\begin{axis}[
width=11em,
height=11em,
grid=both,
grid style={line width=.1pt, draw=gray!10},
major grid style={line width=.2pt,draw=gray!40},
minor tick num=1,
xlabel={$Q$},
ylabel={$P_{\rm succ}$},
x grid style={darkgray176},
xmin=1.0, xmax=1.5,
xtick style={color=black},
y grid style={darkgray176},
ymin=0.0, ymax=1,
ytick style={color=black},
legend style={draw=white!15!black, font=\scriptsize, at={(0.99, 0.98)}, anchor=north east, /tikz/every even column/.append style={column sep=0.2em}},
legend columns=1
]
\addplot [ultra thick, color7] table {qoai_plots/prob_Q_fc1600000000.00_alpha0.35.dat};
\addplot [ultra thick, color2] table {qoai_plots/2_prob_Q_fc1600000000.00_alpha0.35.dat};

\end{axis}
\end{tikzpicture}
\pgfplotsset{scaled y ticks=false}
\begin{tikzpicture}

\definecolor{darkgray176}{RGB}{176,176,176}

\definecolor{crimson2143940}{RGB}{240, 249, 33}
\definecolor{darkorange25512714}{RGB}{253,180,47}
\definecolor{forestgreen4416044}{RGB}{237,121,83}
\definecolor{mediumpurple148103189}{RGB}{204,71,120}
\definecolor{orchid227119194}{RGB}{156, 23, 158}
\definecolor{sienna1408675}{RGB}{92,1,166}
\definecolor{steelblue31119180}{RGB}{13,8,135}

\begin{axis}[
width=11em,
height=11em,
grid=both,
grid style={line width=.1pt, draw=gray!10},
major grid style={line width=.2pt,draw=gray!40},
minor tick num=1,
xlabel={$Q$},
ylabel={$E_{c}+E_{\mathrm{tx}}$ [J]},
x grid style={darkgray176},
xmin=1.0, xmax=1.5,
xtick style={color=black},
y grid style={darkgray176},
ymin=0.06, ymax=0.09,
ytick={0.06,0.07,0.08,0.09},
yticklabels={0.06,0.07,0.08,0.09},
ytick style={color=black},
legend style={draw=white!15!black, font=\scriptsize, at={(0.99, 0.98)}, anchor=north east, /tikz/every even column/.append style={column sep=0.2em}},
legend columns=1
]
\addplot [ultra thick, color2] table {qaoi_plots_new/2_energy_Q_fc1600000000.00_alpha0.35.dat};
\addplot [ultra thick, color7] table {qaoi_plots_new/energy_Q_fc1600000000.00_alpha0.35.dat};

\end{axis}
\end{tikzpicture}
    \caption{Overall success probability and total energy consumption for different values of $Q$, $f_c=1.6$ [GHz] and $\alpha=0.35$.}
    \label{fig:q_prob}
    \vspace{-0.4cm}
\end{figure}

\AM{The role of $T$ is also evident: with $T = 0.4$ s, the reliability threshold $\varrho$ is exceeded only when $Q \gtrsim 1.31$. Relaxing $T$ increases the available transmission time, thereby enabling successful transmission even for lower $Q$. As a consequence, with $T = 0.5$ s, the feasible range of $Q$ that satisfies the reliability constraint broadens significantly. In terms of energy, only the compression energy $E_{\textrm{comp}}$ depends on $Q$, and it grows superlinearly with $Q$ due to the rising computational cost of stronger compression. Thus, energy consumption increases monotonically with $Q$, even in the regime where $P_{\rm succ}$ is maximized. As in Fig.~\ref{fig:alpha_prob}, increasing $T$ alone does not reduce energy consumption: while a larger $T$ broadens the feasible range of $Q$ values that ensure reliability, true energy minimization requires joint optimization of $Q$ together with the other system parameters.} 

\AM{Finally, Fig.~\ref{fig:results_scen2} presents the Pareto front in terms of total energy consumption and \gls{qaoi} for different reliability levels \( \varrho \). Each curve is obtained by finding, for a fixed  \( T \), the minimum total energy achievable across all feasible configurations. Equivalently, each point on the front represents an optimal pairing of \gls{cpu} frequency \( f_c \), compression ratio \( Q \) and $\alpha$ that meets both the \gls{qaoi} and reliability requirements. At all reliability levels, relaxing $T$ or the \gls{qaoi} target substantially lowers the total required energy.}
We also provide the equivalent result under a fast fading channel, which is shown in Fig.~\ref{fig:results_scen2_ff}. As for the time-constrained problem, the difference between the two cases is too small to be appreciable in the plots, and the results and trends in the slow fading scenario can be safely generalized.

\begin{figure}[t]
\begin{tikzpicture}

\definecolor{darkgray176}{RGB}{176,176,176}
\definecolor{steelblue31119180}{RGB}{31,119,180}
\definecolor{crimson2143940}{RGB}{214,39,40}
\definecolor{darkgray176}{RGB}{176,176,176}
\definecolor{darkorange25512714}{RGB}{255,127,14}
\definecolor{forestgreen4416044}{RGB}{44,160,44}
\definecolor{mediumpurple148103189}{RGB}{148,103,189}
\definecolor{steelblue31119180}{RGB}{31,119,180}

\begin{axis}[
width=25em,
height=15.5em,
grid=both,
grid style={line width=.1pt, draw=gray!10},
major grid style={line width=.2pt,draw=gray!40},
minor tick num=1,
xlabel={QAoI [s]},
ylabel={Energy [J]},
unbounded coords=jump,
x grid style={darkgray176},
xmin=0.25, xmax=0.55,
xtick style={color=black},
y grid style={darkgray176},
ymin=0.06, ymax=0.12,
ytick style={color=black},
ytick={0.06, 0.08, 0.1, 0.12},
yticklabels={0.06, 0.08, 0.1, 0.12},
legend style={draw=white!15!black, font=\scriptsize, at={(0.99, 0.98)}, anchor=north east, /tikz/every even column/.append style={column sep=0.2em}},
legend columns=1
]

\addplot [ultra thick, color2]
table[
    x=Quantile,
    y=Energy,
    col sep=comma
] {figures/fast_fading/second_scenario/pareto_rho_0.9.csv};

\addplot [ultra thick, color3]
table[
    x=Quantile,
    y=Energy,
    col sep=comma
] {figures/fast_fading/second_scenario/pareto_rho_0.99.csv};

\addplot [ultra thick, color4]
table[
    x=Quantile,
    y=Energy,
    col sep=comma
] {figures/fast_fading/second_scenario/pareto_rho_0.999.csv};

\addplot [ultra thick, color5]
table[
    x=Quantile,
    y=Energy,
    col sep=comma
] {figures/fast_fading/second_scenario/paret_rho_0.9999.csv};

\addplot [ultra thick, color6]
table[
    x=Quantile,
    y=Energy,
    col sep=comma
] {figures/fast_fading/second_scenario/pareto_rho_0.99999.csv};

\addplot [semithick, color7, only marks, mark=*, mark size=2, mark options={solid}]
table {%
0.390 0.0858
};

\addlegendimage{only marks, thick,  mark=triangle}

\addplot [black, dashed, name path=low] table {%
0.390 0.0858
0.7   0.0858
};
\addplot [black, name path=high] table {%
0.390 0.2
0.7   0.2
};
\addplot [black, dashed, name path=left] table {%
0.390 0.0858
0.390 0.2
};
\addplot[
        fill=black!5,
        draw=none,
        pattern=north west lines,
        pattern color=black
    ] fill between[
        of=high and low
    ];


\legend{$\varrho=0.9$,$\varrho=0.99$, $\varrho=0.999$, $\varrho=0.9999$,  $\varrho=0.99999$, No compression}
\end{axis}

\end{tikzpicture}
    \vspace{-0.5cm}
    \caption{Pareto front for the time-constrained scenario.}\vspace{-0.4cm}
    \label{fig:results_scen2}
\end{figure}
\begin{figure}[t]
\begin{tikzpicture}

\definecolor{darkgray176}{RGB}{176,176,176}
\definecolor{steelblue31119180}{RGB}{31,119,180}
\definecolor{crimson2143940}{RGB}{214,39,40}
\definecolor{darkgray176}{RGB}{176,176,176}
\definecolor{darkorange25512714}{RGB}{255,127,14}
\definecolor{forestgreen4416044}{RGB}{44,160,44}
\definecolor{mediumpurple148103189}{RGB}{148,103,189}
\definecolor{steelblue31119180}{RGB}{31,119,180}

\begin{axis}[
width=25em,
height=15.5em,
grid=both,
grid style={line width=.1pt, draw=gray!10},
major grid style={line width=.2pt,draw=gray!40},
minor tick num=1,
xlabel={QAoI [s]},
ylabel={Energy [J]},
unbounded coords=jump,
x grid style={darkgray176},
xmin=0.25, xmax=0.55,
xtick style={color=black},
y grid style={darkgray176},
ymin=0.06, ymax=0.12,
ytick style={color=black},
ytick={0.06, 0.08, 0.1, 0.12},
yticklabels={0.06, 0.08, 0.1, 0.12},
legend style={draw=white!15!black, font=\scriptsize, at={(0.99, 0.98)}, anchor=north east, /tikz/every even column/.append style={column sep=0.2em}},
legend columns=1
]

\addplot [ultra thick, color2]
table[
    x=Quantile,
    y=Energy,
    col sep=comma
] {figures/slow_fading/second_scenario/pareto_rho_0.9.txt};

\addplot [ultra thick, color3]
table[
    x=Quantile,
    y=Energy,
    col sep=comma
] {figures/slow_fading/second_scenario/pareto_rho_0.99.txt};

\addplot [ultra thick, color4]
table[
    x=Quantile,
    y=Energy,
    col sep=comma
] {figures/slow_fading/second_scenario/pareto_rho_0.999.txt};

\addplot [ultra thick, color5]
table[
    x=Quantile,
    y=Energy,
    col sep=comma
] {figures/slow_fading/second_scenario/pareto_rho_0.9999.txt};

\addplot [ultra thick, color6]
table[
    x=Quantile,
    y=Energy,
    col sep=comma
] {figures/slow_fading/second_scenario/pareto_rho_0.99999.txt};

\addplot [semithick, color7, only marks, mark=*, mark size=2, mark options={solid}]
table {%
0.390 0.0858
};

\addlegendimage{only marks, thick,  mark=triangle}

\addplot [black, dashed, name path=low] table {%
0.390 0.0858
0.7   0.0858
};
\addplot [black, name path=high] table {%
0.390 0.2
0.7   0.2
};
\addplot [black, dashed, name path=left] table {%
0.390 0.0858
0.390 0.2
};
\addplot[
        fill=black!5,
        draw=none,
        pattern=north west lines,
        pattern color=black
    ] fill between[
        of=high and low
    ];


\legend{$\varrho=0.9$,$\varrho=0.99$, $\varrho=0.999$, $\varrho=0.9999$,  $\varrho=0.99999$, No compression}
\end{axis}

\end{tikzpicture}
    \vspace{-0.5cm}
    \caption{Pareto front for the time-constrained scenario under a fast fading channel.}\vspace{-0.3cm}
    \label{fig:results_scen2_ff}
\end{figure}

This highlights the benefit of jointly optimizing compression and communication: relaxing freshness constraints allows lower operating frequencies and higher compression ratios, reducing both computation and transmission energy. The largest gains occur in the low-\gls{qaoi} region, where small freshness trade-offs yield significant \gls{ee} improvements. The trends in Fig.~\ref{fig:results_scen2} resemble the energy-constrained case (Fig.~\ref{fig:results_scen1}), but the time-constrained setting imposes stricter phase separation and tighter timing to meet the reliability target \( \varrho \). Nevertheless, joint optimization achieves substantially lower energy consumption, making it well suited for reliability-critical systems. 

\subsection{Relationship with data size $D$}

\begin{figure}
    \centering
\begin{tikzpicture}

\definecolor{darkgray176}{RGB}{176,176,176}
\definecolor{steelblue31119180}{RGB}{31,119,180}
\definecolor{crimson2143940}{RGB}{214,39,40}
\definecolor{darkgray176}{RGB}{176,176,176}
\definecolor{darkorange25512714}{RGB}{255,127,14}
\definecolor{forestgreen4416044}{RGB}{44,160,44}
\definecolor{mediumpurple148103189}{RGB}{148,103,189}
\definecolor{steelblue31119180}{RGB}{31,119,180}

\begin{axis}[
width=25em,
height=15.5em,
grid=both,
grid style={line width=.1pt, draw=gray!10},
major grid style={line width=.2pt,draw=gray!40},
minor tick num=1,
xlabel={Latency [s]},
ylabel={Energy [J]},
unbounded coords=jump,
x grid style={darkgray176},
xmin=0.2, xmax=0.4,
xtick style={color=black},
y grid style={darkgray176},
ymin=0.05, ymax=0.11,
ytick style={color=black},
ytick={0.04,0.06, 0.08, 0.1, 0.12},
yticklabels={0.04,0.06, 0.08, 0.1, 0.12},
ytick style={color=black},
legend style={draw=white!15!black, font=\scriptsize, at={(0.99, 0.98)}, anchor=north east, /tikz/every even column/.append style={column sep=0.2em}},
legend columns=1
]

\addplot [ultra thick, color2]
table {%
0.257838223530357 0.05363636363636364
0.2500655762508844 0.05454545454545455
0.24514907644333253 0.05545454545454546
0.24168154212674994 0.056363636363636366
0.23906747647525764 0.057272727272727274
0.23700587668068654 0.05818181818181818
0.23532606786841978 0.0590909090909091
0.23392310689318127 0.060000000000000005
0.23272844904386034 0.06090909090909091
0.23169517520692826 0.06181818181818182
0.23078994280032974 0.06272727272727273
0.22998831940973252 0.06363636363636364
0.229271937981215 0.06454545454545454
0.22862668877568276 0.06545454545454546
0.2280415292331048 0.06636363636363637
0.22750767674445346 0.06727272727272728
0.227018046766586 0.06818181818181819
0.22656685275998265 0.06909090909090909
0.22614931560600646 0.07
};

\addplot [ultra thick, color3]
table {%
0.31939716758521375 0.057272727272727274
0.3082191494752853 0.05818181818181818
0.2992800143102894 0.0590909090909091
0.2920166884445029 0.060000000000000005
0.2861243882661245 0.06090909090909091
0.28134564724106514 0.06181818181818182
0.27745855802713887 0.06272727272727273
0.2742853887113007 0.06363636363636364
0.2716663200469764 0.06454545454545454
0.2694717771085173 0.06545454545454546
0.2676072699736741 0.06636363636363637
0.2659975877156022 0.06727272727272728
0.2645895166131076 0.06818181818181819
0.2633362225277832 0.06909090909090909
0.26221297700924673 0.07
0.2611955645274068 0.07090909090909092
0.2602640221045097 0.07181818181818182
0.2594086719919412 0.07272727272727274
0.2586136320188751 0.07363636363636364
0.2578761608991238 0.07454545454545455
0.2571825045137034 0.07545454545454547
0.25653480146367386 0.07636363636363637
0.25591894997072717 0.07727272727272727
0.2553413810198446 0.07818181818181819
0.25478770530561895 0.0790909090909091
0.2542647041641379 0.08
0.253763821543901 0.08090909090909092
0.25328509087836226 0.08181818181818182
0.2528291146467051 0.08272727272727273
0.2523868886837008 0.08363636363636365
0.25196552155817487 0.08454545454545456
0.251558291320677 0.08545454545454546
0.2511647343042478 0.08636363636363636
0.2507902208368653 0.08727272727272728
0.25048484078388505 0.0881818181818182
0.2502648073981755 0.0890909090909091
0.24999638168204713 0.09000000000000001
0.24990590284988035 0.09090909090909091
0.24970341618697667 0.09181818181818183
0.2497465834196203 0.09272727272727274
0.2496113213171722 0.09363636363636364
0.2494999042599545 0.09454545454545454
0.24972664963476188 0.09545454545454546
0.24968465447848706 0.09636363636363637
0.24966702987650205 0.09727272727272729
0.2496738806653149 0.09818181818181819
0.2497053383980839 0.09909090909090909
};

\addplot [ultra thick, color4]
table {%
0.37182912365807264 0.062121212121212126
0.34901017749712526 0.06313131313131314
0.335164113779994 0.06414141414141414
0.3247422365344312 0.06515151515151515
0.31634265901789704 0.06616161616161617
0.3096212874862495 0.06717171717171717
0.30430932829046986 0.06818181818181818
0.30011003813968584 0.0691919191919192
0.2968004423551807 0.07020202020202021
0.2941577671784557 0.07121212121212121
0.2920078322612524 0.07222222222222223
0.29022045484573405 0.07323232323232323
0.2886971783239336 0.07424242424242425
0.28737253725109835 0.07525252525252525
0.2862047119983673 0.07626262626262627
0.28515637239857783 0.07727272727272727
0.28420705577099087 0.07828282828282829
0.28333970906883504 0.0792929292929293
0.28253737275468305 0.0803030303030303
0.2817964457618856 0.08131313131313131
0.28110002122803046 0.08232323232323233
0.28045164972395087 0.08333333333333334
0.2798358845598629 0.08434343434343435
0.27925775243393275 0.08535353535353535
0.27870629495863614 0.08636363636363636
0.27818299232941646 0.08737373737373738
0.27768482072679385 0.08838383838383838
0.2772053008235938 0.08939393939393939
0.2767515104332489 0.0904040404040404
0.27630884393706984 0.09141414141414142
0.27589643623725135 0.09242424242424244
0.2755137331137171 0.09343434343434344
0.2752509610108957 0.09444444444444444
0.2749500749969592 0.09545454545454546
0.2748540393161585 0.09646464646464648
0.2746362886322831 0.09747474747474748
0.27444795005699735 0.09848484848484848
0.27457998363385683 0.0994949494949495
0.2744772170084052 0.10050505050505051
0.2744043897339987 0.10151515151515152
0.274361597511716 0.10252525252525252
0.2743489744569158 0.10353535353535354
};

\addplot [ultra thick, color5]
table {%
0.3962932109152042 0.06545454545454546
0.3685762473592971 0.06636363636363637
0.35275541560143703 0.06727272727272728
0.34158257779481793 0.06818181818181819
0.3324585207489009 0.06909090909090909
0.3249825389524373 0.07
0.318950934453552 0.07090909090909092
0.31413927644102596 0.07181818181818182
0.3103769926970648 0.07272727272727274
0.3074506063586363 0.07363636363636364
0.30517259048084866 0.07454545454545455
0.30336240377878493 0.07545454545454547
0.30187911762207426 0.07636363636363637
0.3006287149517202 0.07727272727272727
0.29954148701150346 0.07818181818181819
0.29858251109537637 0.0790909090909091
0.2977224513404183 0.08
0.2969366939815206 0.08090909090909092
0.2962216926847795 0.08181818181818182
0.2955510711595157 0.08272727272727273
0.2949356229799482 0.08363636363636365
0.2943483620103491 0.08454545454545456
0.29380485921435956 0.08545454545454546
0.2932822104997887 0.08636363636363636
0.2927927521126962 0.08727272727272728
0.2923230223520805 0.0881818181818182
0.29187539370770554 0.0890909090909091
0.29145097961561817 0.09000000000000001
0.2910365020744588 0.09090909090909091
0.2906442855897274 0.09181818181818183
0.29026460351425454 0.09272727272727274
0.28998271062577263 0.09363636363636364
0.2896970493193884 0.09454545454545454
0.2894440966346622 0.09545454545454546
0.2892238974796561 0.09636363636363637
0.28903655050154786 0.09727272727272729
0.2888822012599051 0.09818181818181819
0.2887610367931722 0.09909090909090909
0.28867328126836667 0.1
0.2886191924809274 0.10090909090909092
0.28859905902773536 0.10181818181818184
};

\addplot [ultra thick, color6]
table {%
0.39708237029422966 0.06818181818181819
0.37030388891764615 0.06909090909090909
0.35458202788717524 0.07
0.3430836909197888 0.07090909090909092
0.33391367823672774 0.07181818181818182
0.3265039579873985 0.07272727272727274
0.3206617693159353 0.07363636363636364
0.3161904698087145 0.07454545454545455
0.3129443470375702 0.07545454545454547
0.3106752155679862 0.07636363636363637
0.30904463759311557 0.07727272727272727
0.307794169629008 0.07818181818181819
0.30676215590981776 0.0790909090909091
0.305879947971359 0.08
0.3051060791752408 0.08090909090909092
0.30440589826011233 0.08181818181818182
0.30376599850648695 0.08272727272727273
0.30318456290593654 0.08363636363636365
0.3026318343840325 0.08454545454545456
0.302124702914898 0.08545454545454546
0.3016391061361591 0.08636363636363636
0.3011887438977444 0.08727272727272728
0.3007515332210361 0.0881818181818182
0.3003456457034866 0.0890909090909091
0.29994708504147105 0.09000000000000001
0.29958463176834155 0.09090909090909091
0.29925822699167226 0.09181818181818183
0.29896790528213757 0.09272727272727274
0.29871377598411364 0.09363636363636364
0.2984960096800265 0.09454545454545454
0.29831482824144545 0.09545454545454546
0.2981704974230631 0.09636363636363637
0.2980633212900759 0.09727272727272729
0.2979936379879287 0.09818181818181819
0.2979618165089178 0.09909090909090909
0.29796825420887413 0.1
};

\legend{$D=0.1$~MB, $D=0.25$~MB,  $D=0.5$~MB, $D=0.75$~MB, $D=1$~MB}
\end{axis}

\end{tikzpicture}
    \caption{Pareto region for $\varrho=0.99$ for different data sizes $D$.}
    \label{fig:variable_data_size}
\end{figure}

We have also evaluated the effect of data block size on performance, showing that similar trade-offs appear for values of $D$ different from the one considered in all the previous analysis. In Fig.~\ref{fig:variable_data_size}, we show the Pareto front for the time-constrained problem, considering a fixed reliability value and data sizes $D$ ranging from $100$~kB to $1$~MB. It is noticeable that, for size as small as $100$~kB, the range of the suitable values for the optimization parameters  shrinks and we observe smaller gains in term of latency and energy.

\section{Conclusion}\label{sec:conclusion}
 \AM{This work demonstrated that substantial energy savings in wireless communication systems can be achieved by relaxing strict latency and reliability constraints, without compromising overall system performance. Using a cross-layer optimization approach, we systematically explored the trade-offs between data compression, latency, and energy consumption. Our analysis considered two representative scenarios: energy-constrained systems and time-constrained systems, each reflecting practical deployment contexts such as battery-limited sensor nodes or systems with strict scheduling requirements. Through a detailed end-to-end formulation for both scenarios, we provided a comprehensive view of how system parameters, including \gls{cpu} frequency, compression ratio, and time allocation, can be jointly tuned to minimize energy consumption while meeting application-specific requirements for data freshness and delivery success. In the energy-constrained case, we focused on latency as the key performance metric. In the time-constrained case, we used \gls{qaoi} as a meaningful metric to capture both temporal relevance and reliability under variable compression and transmission dynamics. Simulation results validated the analytical findings and underscored the practical relevance of our framework. In particular, they showed that modest relaxations in reliability and latency requirements can enable significant reductions in energy use, which is a critical outcome for battery-powered or energy-harvesting systems operating under long-term deployment constraints.}

Future work could extend this framework by incorporating lossy compression and a \gls{voi} perspective, where data utility depends on both freshness and reconstruction distortion. This would enable adaptive, application-aware communication strategies that allocate resources based on data importance. Further research may also explore real-time adaptation, multi-hop extensions, and learning-based optimization to improve scalability and robustness in dynamic network environments. Extending the framework to jointly optimize lower-layer communication parameters, including
the transmit power, bandwidth, and blocklength, would enlarge the design space and is an interesting
direction for future work.

\appendices
\section{Convexity of the Compression Latency Quantile}
\label{appendix:convexity}

We now prove that the quantile of the Gamma distribution representing the compression latency is convex under the condition in~\eqref{eq:cond}. It is known that the $\varrho$-quantile of the Gamma distribution of parameters $(\alpha,\beta)$
can be expressed as 
\begin{equation}
    \gamma^{-1}(\alpha,\Gamma(\alpha)\varrho) \cdot \beta,
\end{equation}
where $\gamma$ and $\Gamma$ are the lower incomplete Gamma function and the Gamma function, respectively. 
%
In our case, $\beta$ is the only parameter that depends on $Q$, through the relation
\begin{equation}\label{eq:betaapp}
    \beta = \frac{e^{\psi Q} - e^{\psi}}{\kappa f_c^*(Q)} = K \sqrt{\frac{(e^{\psi Q} - e^{\psi})^3}{E_{\mathrm{max}} - \frac{C}{Q}}},
\end{equation}
where $K$ and $C=\frac{D}{(1-\varepsilon)\eta(\varepsilon)}$ are positive constants which do not depend on $Q$. We are left to show that $\beta$ is convex with respect to $Q$. 
\AZ{To this end, we analyze the sign of the second order derivative of \eqref{eq:betaapp} with respect to $Q$, which is equal to 
\begin{equation}
\label{eq:second_order}
\begin{split}
    \frac{d^2\beta(Q)}{dQ^2} = \frac{1}{2} K \beta(Q) \bigg\{ \frac{1}{2}\bigg[ \frac{C}{Q(QE_{\mathrm{max}}-C)} - \frac{3\psi e^{\psi Q}}{e^{\psi Q} - e^\psi} \bigg]^2 \\ + \frac{3\psi^2 e^\psi e^{\psi Q}}{(e^{\psi Q} - e^{\psi})^2} + \frac{-2QE_{\mathrm{max}} + C}{Q^2E_{\mathrm{max}}-CQ} \bigg\}.
\end{split}    
\end{equation}
}
This function is positive (and, hence, $\beta$ is convex in $Q$), if
\begin{equation}
\label{eq:final_condition}
    \frac{3\psi^2 e^\psi e^{\psi Q}}{(e^{\psi Q} - e^{\psi})^2} + \frac{-2QE_{\mathrm{max}} + C}{Q^2E_{\mathrm{max}}-CQ} \geq 0,
\end{equation}
as all the other terms are non-negative. However, showing that the condition ~\eqref{eq:final_condition} is satisfied is not trivial, as the first two terms have opposite signs. We then give
\begin{align}
    f_c^* &= \sqrt{\frac{f_{\mathrm{max}^3}}{DP_{\mathrm{s,max}}} \frac{E_{\mathrm{max}} - \frac{C}{Q}}{e^{\psi Q} - e^{\psi}}} \\
    \label{eq:first}
    e^{\psi Q} &= e^{\psi} + \frac{f_\mathrm{max}^3}{{f^*_c}^2DP_{\mathrm{s,max}}}\bigg( E_{\mathrm{max}} - \frac{C}{Q}\bigg) \\
    \label{eq:second}
    Q &= \psi^{-1}\log\left[{\frac{f_\mathrm{max}^3}{{f^*_c}^2DP_{\mathrm{s,max}}}\bigg( E_{\mathrm{max}} - \frac{C}{Q}\bigg)} +e^\psi\right].
\end{align}
Substituting \eqref{eq:first} and \eqref{eq:second} into the first term of \eqref{eq:final_condition} and further simplifying, we obtain the final condition 
\begin{equation}
    Q^23\psi^2e^\psi -2QE_{\mathrm{max}} + C \geq 0,
\end{equation}
which is the condition in the theorem statement.

\bibliographystyle{IEEEtran}
\bibliography{biblio}

@article{ts38321,
    author = {3GPP},
    title = {{NR Medium Access Control (MAC)} protocol specification},
    journal = {TS 38.321},
    year = {2026}
}

@ARTICLE{tavildar2006approximately,
  author={Tavildar, S. and Viswanath, P.},
  journal={IEEE Transactions on Information Theory}, 
  title={Approximately universal codes over slow-fading channels}, 
  year={2006},
  volume={52},
  number={7},
  pages={3233-3258}
}

@ARTICLE{ozdogan2019massive,
  author={\"{O}zdogan, \"{O}zgecan and Bj\"{o}rnson, Emil and Larsson, Erik G.},
  journal={IEEE Transactions on Communications}, 
  title={Massive {MIMO} With Spatially Correlated {Rician} Fading Channels}, 
  year={2019},
  volume={67},
  number={5},
  pages={3234-3250}
}

@article{tang2020minimizing,
  title={Minimizing age of information with power constraints: Multi-user opportunistic scheduling in multi-state time-varying channels},
  author={Tang, Haoyue and Wang, Jintao and Song, Linqi and Song, Jian},
  journal={IEEE Journal on Selected Areas in Communications},
  volume={38},
  number={5},
  pages={854-868},
  year={2020},
  publisher={IEEE}
}

@article{hu2022age,
  title={On the age of information and energy efficiency in cellular {IoT} networks with data compression},
  author={Hu, Haonan and Dong, Ying and Jiang, Yan and Chen, Qianbin and Zhang, Jie},
  journal={IEEE Internet of Things Journal},
  volume={10},
  number={6},
  pages={5226-5239},
  year={2022},
  publisher={IEEE}
}

@article{das2025analysis,
  title={Analysis of Age-Energy Trade-off in {IoT} Networks Using Stochastic Geometry},
  author={Das, Songita and Ghatak, Gourab},
  journal={IEEE Transactions on Green Communications and Networking},
  year={2025},
  volume={10},
  pages={236-248},
  publisher={IEEE}
}

@article{huang2021age,
  title={Age-energy tradeoff optimization for packet delivery in fading channels},
  author={Huang, Haitao and Qiao, Deli and Gursoy, M Cenk},
  journal={IEEE Transactions on Wireless Communications},
  volume={21},
  number={1},
  pages={179-190},
  year={2021},
  publisher={IEEE}
}

@article{celebi2021multi,
  title={A multi-objective optimization framework for {URLLC} with decoding complexity constraints},
  author={Celebi, Hasan Basri and Pitarokoilis, Antonios and Skoglund, Mikael},
  journal={IEEE Transactions on Wireless Communications},
  volume={21},
  number={4},
  pages={2786-2798},
  year={2022},
  publisher={IEEE}
}

@article{bisio2024shm,
  title={{SHM} with low-cost, low-energy, and low-rate {IoT} devices: Reducing transmission burden with compressive sensing},
  author={Bisio, Igor and Garibotto, Chiara and Grattarola, Aldo and Lavagetto, Fabio and Sciarrone, Andrea and Zerbino, Matteo},
  journal={IEEE Internet of Things Journal},
  volume={11},
  number={13},
  pages={24323-24333},
  year={2024},
  publisher={IEEE}
}

@inproceedings{zeng2023parallelizing,
  title={Parallelizing stream compression for {IoT }applications on asymmetric multicores},
  author={Zeng, Xianzhi and Zhang, Shuhao},
  booktitle={39th International Conference on Data Engineering (ICDE)},
  pages={950-964},
  year={2023},
  organization={IEEE}
}

@article{elgarhy2024energy,
  title={Energy efficiency and latency optimization for {IoT URLLC and mMTC} use cases},
  author={Elgarhy, Osama and Reggiani, Luca and Alam, Muhammad Mahtab and Zoha, Ahmed and Ahmad, Rizwan and Kuusik, Alar},
  journal={IEEE Access},
  volume={12},
  pages={23132-23148},
  year={2024},
  publisher={IEEE}
}

@article{zhang2022dependent,
  title={Dependent task offloading with energy-latency tradeoff in mobile edge computing},
  author={Zhang, Yanfang and Chen, Jian and Zhou, Yuchen and Yang, Long and He, Bingtao and Yang, Yijin},
  journal={IET Communications},
  volume={16},
  number={17},
  pages={1993-2001},
  year={2022},
  publisher={Wiley Online Library}
}

@article{cong2023latency,
  title={Latency-energy joint optimization for task offloading and resource allocation in {MEC}-assisted vehicular networks},
  author={Cong, Yuliang and Xue, Ke and Wang, Cong and Sun, Wenxi and Sun, Shuxian and Hu, Fengye},
  journal={IEEE Transactions on Vehicular Technology},
  volume={72},
  number={12},
  pages={16369-16381},
  year={2023},
  publisher={IEEE}
}

@article{shao2021communication,
  title={Communication-computation trade-off in resource-constrained edge inference},
  author={Shao, Jiawei and Zhang, Jun},
  journal={IEEE Communications Magazine},
  volume={58},
  number={12},
  pages={20-26},
  year={2021},
  publisher={IEEE}
}

@article{arroyuelo2021engineering,
  title={Engineering Practical {Lempel-Ziv} Tries},
  author={Arroyuelo, Diego and C{\'a}novas, Rodrigo and Fischer, Johannes and K{\"o}ppl, Dominik and L{\"o}bel, Marvin and Navarro, Gonzalo and Raman, Rajeev},
  journal={Journal of Experimental Algorithmics (JEA)},
  volume={26},
  pages={1-47},
  year={2021},
  publisher={ACM New York, NY, USA}
}

@article{alai2013lifetime,
  title={Lifetime dependence modelling using a truncated multivariate gamma distribution},
  author={Alai, Daniel H and Landsman, Zinoviy and Sherris, Michael},
  journal={Insurance: Mathematics and Economics},
  volume={52},
  number={3},
  pages={542-549},
  year={2013},
  publisher={Elsevier}
}

@inproceedings{bagnoli2006log,
  title={Log-concave probability and its applications},
  author={Bagnoli, Mark and Bergstrom, Ted},
  booktitle={Rationality and Equilibrium: A Symposium in Honor of Marcel K. Richter},
  pages={217-241},
  year={2006},
  organization={Springer}
}

@article{ziv1977universal,
  title={A universal algorithm for sequential data compression},
  author={Ziv, Jacob and Lempel, Abraham},
  journal={IEEE Transactions on Information Theory},
  volume={23},
  number={3},
  month=may,
  pages={337-343},
  year={1977},
  publisher={IEEE}
}

@INPROCEEDINGS{Emil@EE,
  author={Björnson, Emil and Larsson, Erik G.},
  booktitle={52nd Asilomar Conference on Signals, Systems, and Computers (SSC)}, 
  title={How Energy-Efficient Can a Wireless Communication System Become?}, 
  year={2018},
  volume={},
  number={},
  pages={1252-1256},
  organization={IEEE},
  doi={10.1109/ACSSC.2018.8645227}}

@article{han2019offloading,
  title={Offloading optimization and bottleneck analysis for mobile cloud computing},
  author={Han, Di and Chen, Wei and Bai, Bo and Fang, Yuguang},  journal={IEEE Transactions on Communications},
  volume={67},
  number={9},
  pages={6153-6167},
  year={2019},
  publisher={IEEE}
}

@article{jovsilo2018selfish,
  title={Selfish decentralized computation offloading for mobile cloud computing in dense wireless networks},
  author={Jo{\v{s}}ilo, Sla{\dj}ana and D{\'a}n, Gy{\"o}rgy},
  journal={IEEE Transactions on Mobile Computing},
  volume={18},
  number={1},
  pages={207-220},
  year={2018},
  publisher={IEEE}
}

@article{li2018wirelessly,
  title={Wirelessly powered crowd sensing: Joint power transfer, sensing, compression, and transmission},
  author={Li, Xiaoyang and You, Changsheng and Andreev, Sergey and Gong, Yi and Huang, Kaibin},
  journal={IEEE Journal on Selected Areas in Communications},
  volume={37},
  number={2},
  pages={391-406},
  year={2018},
  publisher={IEEE}
}

@article{wang2020joint,
  title={Joint optimization of transmission bandwidth allocation and data compression for mobile-edge computing systems},
  author={Wang, Jun-Bo and Zhang, Jinyuexue and Ding, Changfeng and Zhang, Hua and Lin, Min and Wang, Jiangzhou},
  journal={IEEE Communications Letters},
  volume={24},
  number={10},
  pages={2245-2249},
  year={2020},
  publisher={IEEE}
}

@inproceedings{kothiyal2009energy,
  title={Energy and performance evaluation of lossless file data compression on server systems},
  author={Kothiyal, Rachita and Tarasov, Vasily and Sehgal, Priya and Zadok, Erez},
  booktitle={Israeli Experimental Systems Conference (SYSTOR)},
  organization={ACM},
  year={2009}
}

@article{burrows1992line,
  title={On-line data compression in a log-structured file system},
  author={Burrows, Michael and Jerian, Charles and Lampson, Butler and Mann, Timothy},
  journal={ACM SIGPLAN Notices},
  volume={27},
  number={9},
  pages={2-9},
  year={1992},
  publisher={ACM New York, NY, USA}
}

@inproceedings{zhang2020efficient,
  title={Efficient {I/O} for neural network training with compressed data},
  author={Zhang, Zhao and Huang, Lei and Pauloski, J. Gregory and Foster, Ian T.},
  booktitle={34th International Parallel and Distributed Processing Symposium (IPDPS)},
  pages={409-418},
  year={2020},
  organization={IEEE}
}

@article{rizzo1997effective,
author = {Rizzo, Luigi},
title = {Effective erasure codes for reliable computer communication protocols},
year = {1997},
volume = {27},
number = {2},
journal = {ACM SIGCOMM Computer Communication Review},
pages = {24–36},
}

@ARTICLE{Federico@QAoI,
  author={Chiariotti, Federico and Holm, Josefine and Kal{\o}r, Anders E and Soret, Beatriz and Jensen, S{\o}ren K and Pedersen, Torben B and Popovski, Petar},
  journal={IEEE Transactions on Communications}, 
  title={Query Age of Information: Freshness in Pull-Based Communication}, 
  year={2022},
  volume={70},
  number={3},
  pages={1606-1622},
  doi={10.1109/TCOMM.2022.3141786}}

@ARTICLE{Mao@Freq_Power,
  author={Mao, Yuyi and You, Changsheng and Zhang, Jun and Huang, Kaibin and Letaief, Khaled B.},
  journal={IEEE Communications Surveys \& Tutorials}, 
  title={A Survey on Mobile Edge Computing: The Communication Perspective}, 
  year={2017},
  volume={19},
  number={4},
  pages={2322-2358},
  doi={10.1109/COMST.2017.2745201}}

@ARTICLE{Israel@Power_Freq,
  author={Leyva-Mayorga, Israel and Martinez-Gost, Marc and Moretti, Marco and P{\'e}rez-Neira, Ana and V{\'a}zquez, Miguel {\'A}ngel and Popovski, Petar and Soret, Beatriz},
  journal={IEEE Transactions on Communications}, 
  title={Satellite Edge Computing for Real-Time and Very-High Resolution {Earth} Observation}, 
  year={2023},
  volume={71},
  number={10},
  pages={6180-6194},
  doi={10.1109/TCOMM.2023.3296584}}

@article{suman2023statistical,
  title={Statistical characterization of closed-loop latency at the mobile edge},
  author={Suman, Suraj and Chiariotti, Federico and Stefanovic, {\v{C}}edomir and Do{\v{s}}en, Strahinja and Popovski, Petar},
  journal={IEEE Transctions on Communications},
  volume={71},
  number={7},
  pages={4391-4405},
  year={2023},
  publisher={IEEE}
}

@article{marjani2017big,
  title={{Big IoT} data analytics: architecture, opportunities, and open research challenges},
  author={Marjani, Mohsen and Nasaruddin, Fariza and Gani, Abdullah and Karim, Ahmad and Hashem, Ibrahim Abaker Targio and Siddiqa, Aisha and Yaqoob, Ibrar},
  journal={IEEE Access},
  volume={5},
  pages={5247-5261},
  year={2017},
  publisher={Ieee}
}

@article{varghese2021survey,
  title={A survey on edge performance benchmarking},
  author={Varghese \textit{et al.}, Blesson},
  journal={ACM Computing Surveys (CSUR)},
  volume={54},
  number={3},
  pages={1-33},
  year={2021},
  publisher={ACM New York, NY, USA}
}

@article{capra2019edge,
  title={Edge computing: A survey on the hardware requirements in the internet of things world},
  author={Capra, Maurizio and Peloso, Riccardo and Masera, Guido and Ruo Roch, Massimo and Martina, Maurizio},
  journal={Future Internet},
  volume={11},
  number={4},
  pages={100},
  year={2019},
  publisher={MDPI}
}

@article{li2025unified,
  title={Unified Timing Analysis for Closed-Loop Goal-Oriented Wireless Communication},
  author={Li, Lintao and Kal{\o}r, Anders E and Popovski, Petar and Chen, Wei},
  journal={IEEE Transactions on Wireless Communications},
  volume={24},
  number={4},
  pages={2814--2831},
  year={2025},
  publisher={IEEE}
}

@ARTICLE{Mirza@EL,
  author={Baig, Mirza Uzair and Yu, Lei and Xiong, Zixiang and Høst-Madsen, Anders and Li, Houqiang and Li, Weiping},
  journal={IEEE Transactions on Information Theory}, 
  title={On the Energy-Delay Tradeoff in Streaming Data: Finite Blocklength Analysis}, 
  year={2020},
  volume={66},
  number={3},
  pages={1861-1881},
  doi={10.1109/TIT.2019.2954347}}

@article{li2017fundamental,
  title={A fundamental tradeoff between computation and communication in distributed computing},
  author={Li, Songze and Maddah-Ali, Mohammad Ali and Yu, Qian and Avestimehr, A Salman},
  journal={IEEE Transactions on Information Theory},
  volume={64},
  number={1},
  pages={109-128},
  year={2017},
  publisher={IEEE}
}

@article{ballotta2020computation,
  title={Computation-communication trade-offs and sensor selection in real-time estimation for processing networks},
  author={Ballotta, Luca and Schenato, Luca and Carlone, Luca},
  journal={IEEE Transactions on Network Science and Engineering},
  volume={7},
  number={4},
  pages={2952-2965},
  year={2020},
  publisher={IEEE}
}

@book{tse2005fundamentals,
  title={Fundamentals of wireless communication},
  author={Tse, David and Viswanath, Pramod},
  year={2005},
  publisher={Cambridge University Press}
}

@article{YICK20082292,
title = {Wireless sensor network survey},
journal = {Computer Networks},
volume = {52},
number = {12},
pages = {2292-2330},
year = {2008},
issn = {1389-1286},
doi = {https://doi.org/10.1016/j.comnet.2008.04.002},
author = {Jennifer Yick and Biswanath Mukherjee and Dipak Ghosal}
}

@article{barr@energy,
author = {Barr, Kenneth C. and Asanovi\'{c}, Krste},
title = {Energy-aware lossless data compression},
year = {2006},
issue_date = {August 2006},
publisher = {Association for Computing Machinery},
address = {New York, NY, USA},
volume = {24},
number = {3},
issn = {0734-2071},
doi = {10.1145/1151690.1151692},
journal = {ACM Transactions on Computing Systems},
month = aug,
pages = {250–291},
numpages = {42}
}

@ARTICLE{Time@Petar,
  author={Popovski, Petar and Chiariotti, Federico and Huang, Kaibin and Kalør, Anders E. and Kountouris, Marios and Pappas, Nikolaos and Soret, Beatriz},
  journal={Proceedings of the IEEE}, 
  title={A Perspective on Time Toward Wireless 6G}, 
  year={2022},
  volume={110},
  number={8},
  pages={1116-1146},
  doi={10.1109/JPROC.2022.3190205}}

@ARTICLE{Chen@Tradoff,
  author={Chen, Yan and Zhang, Shunqing and Xu, Shugong and Li, Geoffrey Ye},
  journal={IEEE Communications Magazine}, 
  title={Fundamental trade-offs on green wireless networks}, 
  year={2011},
  volume={49},
  number={6},
  pages={30-37},
  doi={10.1109/MCOM.2011.5783982}}

@ARTICLE{Shao@Comm_Comp,
  author={Shao, Jiawei and Zhang, Jun},
  journal={IEEE Communications Magazine}, 
  title={Communication-Computation Trade-off in Resource-Constrained Edge Inference}, 
  year={2020},
  volume={58},
  number={12},
  pages={20-26},
  doi={10.1109/MCOM.001.2000373}}

@article{long2025energy,
  title={Energy-latency tradeoff for task offloading and resource allocation in vehicular edge computing},
  author={Long, Yuxuan and Wang, Zhenyu and Lan, Shizhan and Zhang, Rui and Xu, Kai},
  journal={Computer Networks},
  volume={258},
  pages={111026},
  year={2025},
  publisher={Elsevier}
}

@ARTICLE{SourceVsChannelCode,
  author={Hochwald, B. and Zeger, K.},
  journal={IEEE Transactions on Information Theory}, 
  title={Tradeoff between source and channel coding}, 
  year={1997},
  volume={43},
  number={5},
  pages={1412-1424},
  doi={10.1109/18.623141}}

@ARTICLE{Hag@SourceControlled_Channel,
  author={Hagenauer, J.},
  journal={IEEE Transactions on Communications}, 
  title={Source-controlled channel decoding}, 
  year={1995},
  volume={43},
  number={9},
  pages={2449-2457},
  doi={10.1109/26.412719}}

@ARTICLE{Avranas@EL_Tradoff,
  author={Avranas, Apostolos and Kountouris, Marios and Ciblat, Philippe},
  journal={IEEE Journal on Selected Areas in Communications}, 
  title={Energy-Latency Tradeoff in Ultra-Reliable Low-Latency Communication With Retransmissions}, 
  year={2018},
  volume={36},
  number={11},
  pages={2475-2485},
  doi={10.1109/JSAC.2018.2874143}}

@ARTICLE{Berry@Delay,
  author={Berry, R.A. and Gallager, R.G.},
  journal={IEEE Transactions on Information Theory}, 
  title={Communication over fading channels with delay constraints}, 
  year={2002},
  volume={48},
  number={5},
  pages={1135-1149},
  doi={10.1109/18.995554}}

@article{burd1996processor,
  title={Processor design for portable systems},
  author={Burd, Thomas D and Brodersen, Robert W},
  journal={Journal of VLSI Signal Processing Systems for Signal, Image and Video Technology},
  volume={13},
  number={2},
  pages={203-221},
  year={1996},
  publisher={Springer}
}

@inproceedings{de2013energy,
  title={The energy/frequency convexity rule: Modeling and experimental validation on mobile devices},
  author={De Vogeleer, Karel and Memmi, Gerard and Jouvelot, Pierre and Coelho, Fabien},
  booktitle={10th International Conference on Parallel Processing and Applied Mathematics (PPAM)},
  pages={793-803},
  year={2013},
  organization={Springer}
}

@article{li2024reliability,
  title={Reliability-latency-rate tradeoff in low-latency communications with finite-blocklength coding},
  author={Li, Lintao and Chen, Wei and Popovski, Petar and Letaief, Khaled B},
  journal={IEEE Transactions on Information Theory},
  year={2025},  
  volume={71},
  number={1},
  pages={360-389},
  publisher={IEEE}
}

@article{tan2015third,
  title={The third-order term in the normal approximation for the {AWGN} channel},
  author={Tan, Vincent Yan Fu and Tomamichel, Marco},
  journal={IEEE Transactions on Information Theory},
  volume={61},
  number={5},
  pages={2430-2438},
  year={2015},
  publisher={IEEE}
}

@article{polyanskiy2010channel,
  title={Channel coding rate in the finite blocklength regime},
  author={Polyanskiy, Yury and Poor, H Vincent and Verd{\'u}, Sergio},
  journal={IEEE Transactions on Information Theory},
  volume={56},
  number={5},
  pages={2307-2359},
  year={2010},
  publisher={IEEE}
}

@article{li2026task,
  title={Task-oriented computation offloading for edge inference: An integrated {Bayesian} optimization and deep reinforcement learning framework},
  author={Li, Xian and Bi, Suzhi and Zhang, Ying-Jun Angela},
  journal={IEEE Transactions on Mobile Computing},
  year={2026}, 
  volume={25},
  number={8},
  pages={13019-13033},
  publisher={IEEE}
}

@book{rappaport2010wireless,
  title={Wireless communications: Principles and practice, 2nd Ed.},
  author={Rappaport, Theodore S},
  year={2010},
  publisher={Prentice Hall}
}

\end{document}